\documentclass[reprint,amsmath,amssymb,aps,pre,floatfix]{revtex4-1}

\usepackage{graphicx}
\usepackage{dcolumn}
\usepackage{bm}
\usepackage[hidelinks]{hyperref}
\usepackage[mathlines]{lineno}
\usepackage{soul}

\hypersetup{
    colorlinks,
    linkcolor={blue},
    citecolor={blue},
    urlcolor={blue}
}

\begin{document}

\title{Spatial characterization of turbulent channel flow via complex networks}

\author{G. Iacobello}
\email{giovanni.iacobello@polito.it}
	\affiliation{Department of Mechanical and Aerospace Engineering, Politecnico di Torino, 10129 Turin, Italy}
\author{S. Scarsoglio}
	\affiliation{Department of Mechanical and Aerospace Engineering, Politecnico di Torino, 10129 Turin, Italy}
\author{J.G.M. Kuerten}
	\affiliation{Department of Mechanical Engineering, Eindhoven University of Technology, P.O. Box 513, 5600 MB Eindhoven, The Netherlands}
\author{L. Ridolfi}
	\affiliation{Department of Environmental, Land and Infrastructure Engineering, Politecnico di Torino, 10129 Turin, Italy}

\date{\today}

	\begin{abstract}

		A network-based analysis of a turbulent channel flow numerically solved at $Re_{\tau}=180$ is proposed as an innovative perspective for the spatial characterization of the flow field. Two spatial networks corresponding to the streamwise and wall-normal velocity components are built, where nodes represent portions of volume of the physical domain. For each network, links are active if the correlation coefficient of the corresponding velocity component between pairs of nodes is sufficiently high, thus unveiling the strongest kinematic relations. Several network measures are studied in order to explore the inter-relations between nodes and their neighbors. Specifically, long-range links are localized between near-wall regions and associated with the temporal persistence of coherent patterns, namely high and low speed streaks. Furthermore, long-range links play a crucial role as intermediary for the kinematic information flow, as emerges from the analysis of indirect connections between nodes. The proposed approach provides a framework to investigate spatial structures of the turbulent dynamics, showing the full potential of complex networks. Although the network analysis is based on the two-point correlation, it is able to advance the level of information, by exploiting the texture created by active links in all directions. Based on the observed findings, the current approach can pave the way for an enhanced spatial interpretation of the turbulence dynamics.
	\end{abstract}

\maketitle

\section{Introduction}

		Turbulence dynamics represents an interdisciplinary branch of research with a wide range of interests, since most flows occurring in nature or in industrial applications are turbulent \cite{tennekes1972first}. Examples are the oceanic currents and the fluvial streams, the atmospheric boundary layer, combustion processes, as well as wake flows behind vehicles, flows through pipes and pumps. Turbulent flows are characterized by complex spatio-temporal fields with many interacting scales, displaying an intrinsic chaotic behaviour \cite{nelkin1992sense, pope}. A wide range of mathematical tools -- e.g., high order moments, structure and correlation functions, spectral and principal component (POD) analysis -- have been largely employed to extract information for a better understanding of turbulence, both theoretical and phenomenological.
	
		Within turbulent flows, particular attention has been given to wall turbulence, mainly due to the importance of the fluid-wall interaction and related energy losses \cite{jimenez2013near}. The main wall turbulence topics include the analysis of the wall-normal structure and scaling (e.g., mean velocity and fluctuations distributions), the investigation and characterization of coherent structures (e.g., near-wall streaks or large scale motions), as well as the interaction between different turbulent scales, that have been fostered by the increasing possibility to explore higher Reynolds number flows \cite{marusic2010wall, smits2011high}. In fact, only in the latest decades experimental and numerical simulations have provided a sufficiently large amount of detailed data, driven by the notable increase of the available computational capabilities \cite{jimenez2013near, jim2003computing}. However, although widely investigated, several issues regarding wall-bounded turbulent flows still remain open, such as detection and characterization of coherent structures as well as their implementation in engineering models \cite{mckeon2007introduction, marusic2010wall, smits2011high, chen2014velocity}. Therefore, new interdisciplinary approaches are required, in order to properly handle the large amount of detail represented by the so called \textit{big-data} \cite{pollard2017whither}.
	
	 A remarkable example of innovative tool for the analysis of real-world complex systems is offered by complex network theory. By combining graph theory and other disciplines such as statistical mechanics and data mining \cite{WattsStrogatz1998, boccaletti2006complex, newman2010networks}, the complex network approach proves to be a powerful and versatile framework, in which a system can be studied through the properties of its constituents  (corresponding to the \textit{nodes} of the network) and the inter-relations (namely, the \textit{links}) between them \cite{albert2002statistical}. From this point of view, complex networks act as a bridge between the graph representation and the underlying complex system \cite{zweig2016network}. Network science has successfully been applied to many research fields \cite{caldarelli2007scale, kaluza2010complex, tsonis2004architecture}, from social dynamics to Internet, economy, climate, biology and transportation systems. The application of the complex network analysis to physical or engineering problems is a very recent research frontier. Specifically, the investigation of fluid flow regimes has mainly comprised geophysical flows \cite{ser2015flow, tupikina2016}, turbulent jets \cite{Shirazi2009, charakopoulos2014}, two-phase flows \cite{gao2009flow, gao2013, gao_ijbc_2017}, fully developed turbulence \cite{liu2010,Manshour2015_fullydev,iacobello2018visibility}, reacting flows \cite{murugesan2015, singh2017network}, isotropic turbulence \cite{taira2016, scarsoglio2016complex}, and biomedical flows \cite{scarsoglio2017time}. In this context, the network-based approach relies on spatio-temporal data, so that networks have been typically built in two ways: (i) by assigning nodes to the samples of time-series at a fixed position, and exploiting the temporal structure of the series to create links \cite{gao2017complex}; (ii) by identifying nodes as spatial locations in the flow field, and using a functional relation between the time-series to activate links \cite{scarsoglio2016complex, lindner2017, donges2009complex}.

	In this work, we propose a network-based analysis of fully-developed turbulent channel flow to offer an innovative approach to study wall turbulence dynamics. To the best of our knowledge, the application of network analysis to spatially investigate wall turbulent data has not been pursued to date. A fully-developed turbulent channel flow was first solved through a direct numerical simulation (DNS), in which the velocity field is computed in each grid point. Two networks are built, corresponding to the streamwise and wall-normal velocity components, respecively. The nodes of the network were associated to the cell volume of the spatial grid points of the simulation, leading to a spatial network where nodes represent physical portions of the domain. For each network, a link between a pair of nodes is activated if the Pearson correlation coefficient of the corresponding velocity component is above a suitable threshold, thus highlighting the strongest (linear) kinematic relations. Although other metrics able to account for nonlinear relations (e.g., mutual information) can be exploited to build the networks, the correlation coefficient -- due to its simplicity and its extensive use in the turbulence literature \cite{pope, wallace2014space, he2017space} -- represents the most suitable metric to show the potential of complex network analysis of wall turbulence.
	
	The present approach provides a framework to systematically investigate the turbulent dynamics (i) by preserving the spatial information, and (ii) by exploiting the topology of the interactions between the components. Differently to other techniques, where the spatial collocation of the two-point correlation is lost due to pre- or post-processing operations (e.g., see Ref. \cite{chen2014velocity}), here the outcomes can be precisely localized in the physical domain and retain, through the network formalism, the multi-point effects of direct and indirect links in all directions.

	The paper is organized as follows. The methodology adopted is reported in Sec. \ref{Sec:methods}. Specifically, the overall features of the simulation are described in \ref{subsec:data}, the network definitions and metrics are introduced in \ref{subsec:definitions}, and the network building procedure is explained in \ref{subsec:net_build}. The results are presented in Sec. \ref{Sec:results}, highlighting the inter-relations between nodes and their neighbors (i.e., between different channel regions)
. The network analysis is carried out at three different scales, namely global scale (Sec.\ref{subsec:global}), meso-scale (Sec.\ref{subsec:meso}), and local scale (Sec.\ref{subsec:local}). Finally, the conclusions are outlined in Sec. \ref{Sec:conclusions}.

\section{Methods}\label{Sec:methods}

\subsection{Data description and pre-processing}\label{subsec:data}
	Direct numerical simulation (DNS) of the turbulent channel flow was solved at $Re_{\tau}=180$, where $Re_{\tau}$ is the Reynolds number based on the frictional velocity, $u_{\tau}$. The geometrical domain has a length $4\pi H$ in the streamwise direction $x$ (with 576 grid points), $2H$ in the wall-normal direction $y$ (with 193 grid points) and $4/3 \pi H$ in the spanwise direction $z$ (with 288 grid points), where $H$ denotes half the channel height. In particular, the computational domain is periodic in the $x$ and $z$ directions. The number of time steps in the simulation is 5000, which corresponds to a time $T u_{\tau}/H=1.25$, or $T^+=225$, where  the superscript $+$ denotes wall units. Time is expressed in units of $H/u_{\tau}$ and velocity in units of $u_{\tau}$. Full details of the simulation are reported in Appendix A.

\noindent In order to have a manageable network size, we reduced the computational domain in the streamwise and spanwise homogeneous directions. Since the results are mainly dependent on the wall-normal coordinate (i.e., the inhomogeneous direction), this operation does not alter the significance of the results. We then selected one out of every four grid points in the $x$ direction, resulting in $N_x'=144$ equally spaced grid points, and $N_z'=150$ consecutive grid points in the $z$-direction. As a result, the streamwise spacing increases as $\Delta x'=4\Delta x$, while the spanwise size of the domain reduces to $L_z'=L_z N_z'/N_z=25/36 \pi H$. In the wall-normal direction, instead, the grid points corresponding to the walls (i.e., $y^+=0$ and $y^+=360$) were excluded, since in those locations the velocity time-series are constantly zero, so that $N_y'=191$. By doing so, the resulting domain is only periodic in the $x$ direction. The final domain size is $(L_x,L_y,L'_z)=(4\pi H,2H,25/36\pi H)$, corresponding to a total volume $V_{tot}=L_x L_y L'_z\simeq 54.8 H^3$, while the final spatial discretization is $(N_x',N_y',N_z')=(144,191,150)$. 

	The whole length of the domain is maintained in the $x$ and $y$ directions because the streamwise and wall-normal directions are of crucial importance. Indeed, they are the directions of advection and inhomogeneity of the flow, respectively. Moreover, taking the whole domain in the streamwise direction guarantees to entirely capture elongated turbulent structures such as streaks, which have scales of the order of $L_x$ \cite{jimenez2013near}. 

	\noindent The choice to take a coarser spatial discretization in the streamwise direction than in the spanwise direction is motivated by the typical spatial scales of the correlation field in the homogeneous directions. Indeed, the spatially-averaged correlation evaluated along the spanwise direction (at fixed $y^+$) decreases more rapidly than the one evaluated in the streamwise direction (see Fig. \ref{fig:Umean_autocorr}b in Appendix A).

\begin{figure}[h]
	\centering
	\includegraphics[width=\linewidth]{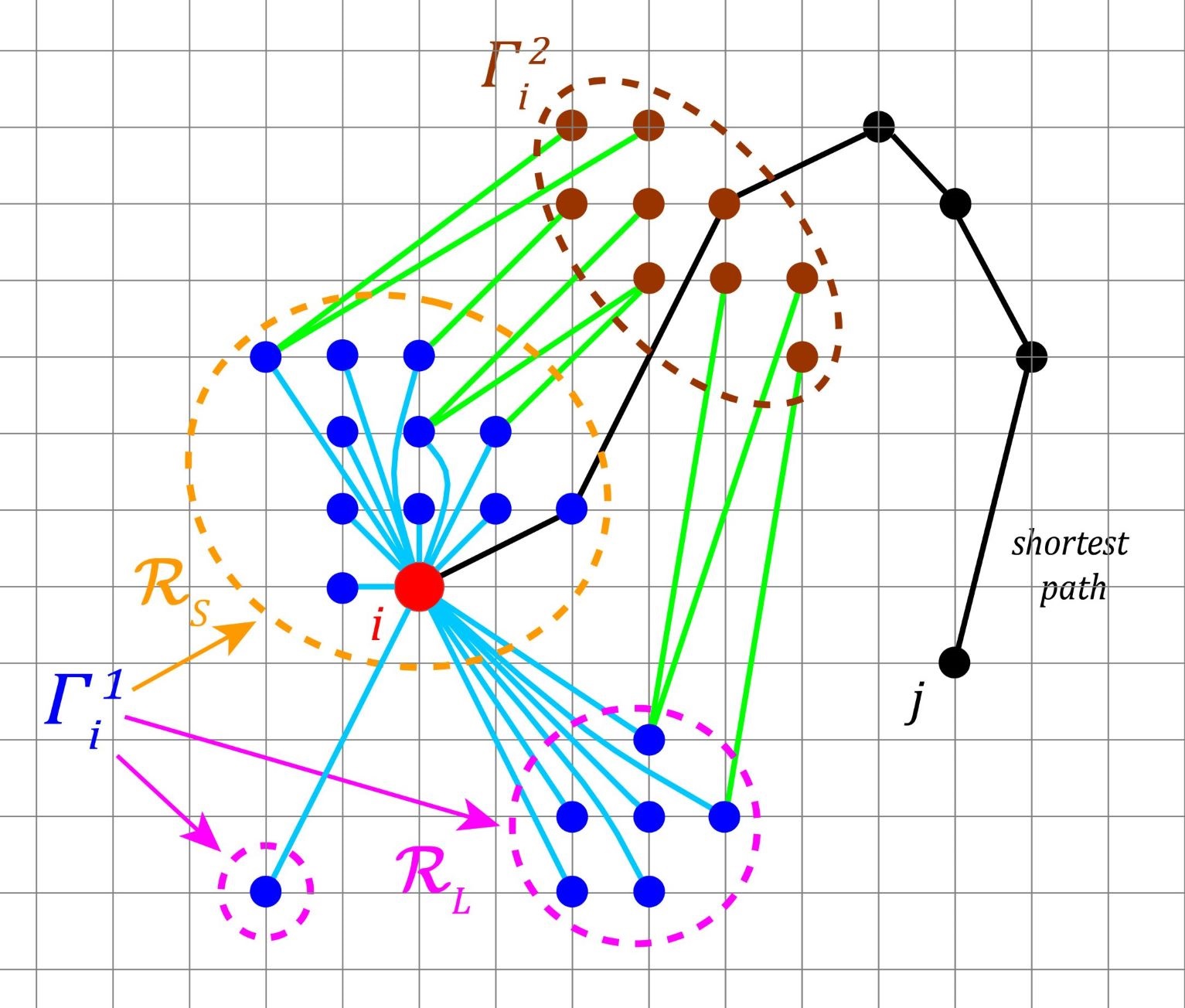}\hfill
	\caption{Sketch of a node, $i$, (depicted in red) and its first and second neighborhood (illustrated as blue and brown dots, respectively). The first neighbors in the short- and long-range regions, $\mathcal{R}_S$ and $\mathcal{R}_L$, are highlighted by dashed circles colored in orange and magenta, respectively. An example of shortest path between node $i$ and a generic node $j$ is shown in black. \label{fig:example_neigb}}
\end{figure}
	
	\subsection{Complex networks: definitions and metrics}\label{subsec:definitions}
		In this section, a summary of the network metrics investigated in the present work is reported. Some of the concepts described here are typically used in network theory (e.g., the degree centrality, the neighborhoods and the shortest path) \cite{boccaletti2006complex, newman2003structure, costa2007characterization}, while some others are here introduced \textit{ad hoc} (e.g., the volume-weighted connectivity and the number of regions).
		
		A network is defined as a graph $G(N_v,N_e) = (\mathcal{V},\mathcal{E})$, where $\mathcal{V}=\lbrace 1,2,...,N_v\rbrace$ is a set of $N_v$ labeled nodes (or vertices) and $\mathcal{E}=\lbrace 1,2,...,N_e\rbrace$ is a set of $N_e$ links (or edges). The structure created by node interactions is called the network topology \cite{zanin2016combining}; a \textit{complex network} is therefore a network with non-trivial topological features.
		
		The \textit{adjacency matrix}, $A_{ij}$, defined as
		\begin{equation} \label{eq:AdjM}
			A_{i,j} = \left \{\begin{array}{l} 0$, if $i=j$ or $\lbrace i,j\rbrace\not\in\mathcal{E},\\ 1$, if $\lbrace i,j\rbrace\in\mathcal{E}, 	
			\end{array}	
			\right.
		\end{equation}

\noindent indicates the existence of a link between a pair $\lbrace i,j\rbrace$ of nodes. In particular, in this study the direction of the link is not taken into account, i.e. $A_{i,j}=A_{j,i}$, and the network is undirected.

		The degree centrality, $k$, of a node $i$ is defined as $k(i)=\sum_{j=1}^{N_v} A_{ij}$, quantifying the number of nodes linked to $i$. The degree centrality is then a measure of the cardinality of the set of nodes directly connected to a node $i$. This set is called the \textit{first neighborhood}, $\Gamma_i^1$, of $i$ and the nodes belonging to $\Gamma_i^1$ are called \textit{first neighbors} of $i$ (e.g., see in Fig. \ref{fig:example_neigb} the red node and its first neighbors in blue). In general, a set of nodes constitutes the $N$-th neighborhood, $\Gamma_i^N$, of a node $i$ if the minimum number of different links connecting $i$ and  $\Gamma_i^N$ is equal to $N$, with $N\geq 0$ and, by definition, $\Gamma_i^0=i$. For example, in Fig. \ref{fig:example_neigb} the second neighborhood of the red node $i$, $\Gamma_i^2$, is shown as brown points. Therefore, the $N$ value indicates the topological distance of the \textit{shortest path} between a node $i$ and the nodes in $\Gamma_i^N$ (see the black path in Fig. \ref{fig:example_neigb}, representing the shortest path between the red node $i$ and the node $j\in\Gamma_i^6$). The \textit{$N$-th cumulative neighborhood}, $\Gamma_i^{N,c}$, of a node $i$ is the union of its first $N$ neighborhoods (including $i$).
		
		A network is made up of a discrete set of nodes, that in this work correspond to fixed spatial positions in the computational domain. An appropriate way to represent a non-uniform physical domain is to assign to each node a weight indicating the spatial extension of that node. This choice is due to the inhomogeneity of the computational grid in the $y^+$ direction, so that nodes at different $y^+$ have different weights. This approach is typically adopted in climate networks, where nodes represent regions of different area on the Earth's surface as a function of the latitude \cite{tsonis2006networks, donges2009complex}. In this work, we assign to each node $i=1,..,N_v$ a weight, $V_i(y^+)=(\Delta x' \Delta y_i(y^+) \Delta z)$, equal to the volume of that node. In particular, in the streamwise and spanwise direction the spacings are uniform (i.e., $\Delta x'=4\Delta x$ and $\Delta z$ are constant), while in the wall-normal direction the spacing depends on $y^+$. The $\Delta y_i(y^+)$ length is then calculated as the sum of the previous and next half-heights of the grid spacing in the $y$ direction, $\Delta y_i=(y_{i+1}-y_{i-1})/2$. Accordingly, we define the \textit{volume-weighted connectivity} of a node $i$ as

		 \begin{equation}\label{eq:VWC}
		 	C^w(i)=\frac{1}{V_{tot}} \sum_{j=1}^{N_v}{A_{ij}^+V_j},
		\end{equation}

		\noindent where $V_{tot}$ is the total volume of the physical domain, and $A_{i,j}^+=A_{i,j}+\delta_{i,j}$ is the extended adjacency matrix \cite{heitzig2012node}, with $\delta_{i,j}$ the Kronecker delta. The extended adjacency matrix is used in Eq. (\ref{eq:VWC}) to ensure that $C^w(i)$ ranges in the interval $\left[0,1\right]$. $C^w(i)$ represents the node-weighted degree of a node $i$ and corresponds to the fraction of volume to which the node $i$ is connected. As the degree centrality, $C^w$ is an indicator of the most important vertices in a network. The fraction of nodes in the network with a given value of $C^w$ is the \textit{$C^w$ distribution}, $p(C^w)$, and represents the probability that a randomly chosen node has a given value of $C^w$. In order to smooth the statistical fluctuations present in the tails of $p(C^w)$ \cite{boccaletti2006complex}, we define the \textit{cumulative $C^w$ distribution} as
		\begin{equation} \label{eq:PVWCcum}
			P(C^w)=1-\sum_{C'^w=0}^{C^w} {p(C'^w)},
		\end{equation}		
	which is the probability to find a node with volume-weighted connectivity greater than or equal to $C^w$.
				
		The \textit{average nearest neighbors $C^w$} of a node $i$ is defined as \cite{heitzig2012node}
		
		\begin{equation} \label{eq:knn_VWC}
			C^w_{nn}(i)=\frac{1}{C^w(i)}\sum_{j\in \Gamma_i^1}{\frac{V_j}{V_{tot}}C^w(j)},
		\end{equation}						
		
		\noindent representing the weighted average of the $C^w$ values of the first neighbors of $i$. If there is no correlation between $C^w(i)$ and $C^w_{nn}(i)$ the network is said \textit{non-assortative}; if, instead, $C^w_{nn}(i)$ is an increasing/decreasing function of $C^w(i)$ the network is classified as \textit{assortative/disassortative}.
		
		Since to each node of the network corresponds a volume in a fixed spatial grid position, nodes that are close in space can be grouped according to a connectivity criterion. In this work, we say that a set of nodes forms a \textit{spatially-connected region} (or simply a \textit{region}), $\mathcal{R}$, if each node in $\mathcal{R}$ is distant one grid spacing (in any Cartesian direction, $\pm\Delta x$, $\pm\Delta y(y^+)$ or $\pm\Delta z$) from at least another node of the set \cite{lozano2012three}. The volume, $V_{\mathcal{R}}$, occupied by a region $\mathcal{R}$, is $V_{\mathcal{R}}=\sum_i{V_i}$, with $i\in\mathcal{R}$. Notice that in our definition the nodes in a region only satisfy a geometrical condition, but they are not necessarily linked with each other (topological condition). Accordingly, it is possible to group the $N$-th neighborhood, $\Gamma_i^N$, of a node $i$ into a number $\mathcal{N}(\Gamma_i^N)$ of spatially-connected regions. For example, in Fig. \ref{fig:example_neigb}, the first neighborhood of the red node $i$ can be partitioned into $\mathcal{N}(\Gamma_i^1)=3$ regions (colored in blue), while its second neighborhood, $\Gamma_i^2$, forms only one region (colored in brown). In particular, we say that a node $j$ is a \textit{short-range neighbor} of $i$ if both $i$ and $j$ belong to the same region, $\mathcal{R}_S$ (see nodes grouped in orange in Fig. \ref{fig:example_neigb}). On the contrary, $j$ is called a \textit{long-range neighbor} of $i$, if $j$ and $i$ do not belong to the same region. The sets of long-range neighbors of a node $i$ are then indicated as $\mathcal{R}_L$ (e.g., see nodes grouped in magenta in Fig. \ref{fig:example_neigb}). For every node in a network, $\mathcal{N}(\mathcal{R}_S)=1$ and $\mathcal{N}(\Gamma^1)=\mathcal{N}(\mathcal{R}_S)+\mathcal{N}(\mathcal{R}_L)=1+\mathcal{N}(\mathcal{R}_L)$, thus long-range neighbors are present only if $\mathcal{N}(\Gamma^1)>1$. By extension, we refer to short- and long-range links to indicate the connections between pairs of short- and long-range neighbors, respectively. It should be pointed out that long-range neighbors of a node $i$ are nodes "detached" from the short-range region, regardless of the physical (Euclidean) distance from $i$. Namely, there is a spatial gap (at least greater than one grid step, in each direction) that divides short- and long-range regions.

		Finally, the \textit{weighted physical distance}, $d_{W,\alpha}(i,j)$, in the Cartesian direction $\alpha\in\left\lbrace x,y,z\right\rbrace$, between a node $i$ and one of its first neighbor $j$ is here defined as $d_{W,\alpha}(i,j)=|\alpha_i-\alpha_j| V_j/V_{tot}$, with $j\in\Gamma_i^1$. The average weighted physical distance between a node $i$ and its first neighbors in a region $\mathcal{R}$ is then evaluated as
		\begin{equation} \label{eq:dist_mean}
			\left\langle d_{W,\alpha}\right\rangle(i)=\frac{1}{V_{\mathcal{R}}}\sum_{j\in \mathcal{R}}{d_{W,\alpha}(i,j)},
		\end{equation}	
		\noindent with $V_{\mathcal{R}}=\sum_{j\in\mathcal{R}}{V_j/V_{tot}}$ and $\mathcal{R}\subseteq\Gamma_i^1$.

	\subsection{Network building}\label{subsec:net_build}
	
		We firstly assigned a node to each selected grid point, resulting in spatial networks with $N_v=144\times191\times150 = 4125600\sim10^6$ nodes. The Pearson correlation coefficients, $C_{i,j}$, based on the time-series of the streamwise and wall-normal velocity components, $u(x,y,z,t)$ and $v(x,y,z,t)$, were evaluated for each pair of nodes, $\left\lbrace i,j\right\rbrace$. The correlation coefficients are calculated from the whole simulation time $T=1.25$ (by taking all the $N_T=5000$ time samples), corresponding to about $1.5$ times the flow through time ($L_x/U_b$, where $U_b$ is the bulk velocity). Links are active if the absolute value of the correlation coefficient is greater than a suitable threshold, $\tau$, which was here set equal to $0.85$, i.e., $|C_{i,j}|>\tau$. A high value of the threshold $\tau$ was chosen to highlight the strongest positive and negative correlations and to have a manageable number of links. Therefore, the total number of links, $N_e$, depends on the correlation threshold value. For $\tau=0.85$, we obtain $N_{e,u}=857693107\sim10^9$ and $N_{e,v}=226842435\sim10^8$ links, for the network based on the $u$ and $v$ components, respectively. The corresponding network edge density values are $\rho_{e,u}=N_{e,u}/N_{e,tot}\approx 10^{-4}$ and $\rho_{e,v}=N_{e,v}/N_{e,tot}\approx 10^{-5}$, where $N_{e,tot}=N_v(N_v-1)/2$ is the maximum number of possible edges in a network of $N_v$ nodes. The values of $\rho_e$ are very low, meaning that the networks are sparse. In general, the choice of the threshold is a non-trivial aspect in the analysis of correlation networks. A threshold that is too high leads to extremely sparse networks, in which mainly trivial connections are unveiled. On the contrary, a too low value of $\tau$ results in networks where the statistical significance of the links is arguable, thus making the interpretation of the network structure confused or misleading. Consequently, to highlight the strongest (linear) relations we performed the main analysis at the same high threshold value (i.e., $\tau=0.85$) for both $u$ and $v$, while a parametric analysis of the results for different $\tau$ values is reported in Appendix C.

		 \noindent The networks so built allow us to spatially characterize the turbulent channel flow from a kinematic point of view (since the velocity components were considered), with linear relations among nodes (since the Pearson correlation was evaluated). Since the continuity and Navier-Stokes equations are numerically solved through a direct numerical simulation, and they represent conservation laws of mass and momentum, the dynamical constraints are actually embedded in the resulting flow field. Therefore, the flow dynamics features and constraints are inherited in the kinematic description of the relations in the flow. Specifically, the streamwise and wall-normal velocities were selected here because they are two of the most significant variables to characterize a turbulent channel flow. Indeed, the streamwise velocity is the component containing the largest part of the kinetic energy, while $v$ is the velocity component corresponding to the inhomogeneous direction \cite{jimenez2018coherent}. However, the procedure carried out in this work can also be applied to other physical quantities (e.g., turbulence kinetic energy, or the vorticity field). The correlation-threshold approach is one of the simplest and most adopted techniques to construct complex networks \cite{tumminello2005tool, aste2006dynamical}, but other inter-node relations are also exploited (e.g., mutual information \cite{donges2009complex, hlinka2014non}, Granger causality \cite{charakopoulos2018dynamics} or eigen-techniques \cite{donges2015complex}). Our choice of the correlation as measure to create links is in line with the exploratory nature of this work. Due to its simplicity and its broad use in the turbulence literature \cite{he2017space, wallace2014space}, correlation represents the most suitable metric to start showing the potential of complex networks applied to wall-bounded turbulence. Although nonlinear effects might be included by exploiting, for instance, the mutual information, its evaluation would require a detailed and refined phase of calibration and testing, which is out of the scope of this work.

\section{Results and Discussion}\label{Sec:results}

Results are presented to highlight how the kinematic information (i.e., related to the $u$ and $v$ velocity components) spatially flows in the temporal window considered, and how this kinematic information is organized at three different scales:
		
	\begin{enumerate}
		\item \textit{Global scale}. The overall characteristics of the whole network are investigated (i.e., considering all nodes, without any distinction); in particular, the centrality of nodes (in terms of $C^w$ probability) and the similarity among nodes (in terms of assortativity, $C^w_{nn}$) are explored.
		\item \textit{Meso-scale}. The attention is given to the topological features of groups of nodes; specifically, we study the network metrics (i) as a function of $y^+$, and (ii) focusing on the most central nodes (in terms of $C^w$).
		\item \textit{Local scale}. The analysis is focused on single nodes; here, we explore the neighborhoods of representative nodes with extremely different features.
	\end{enumerate}

	The analysis at different scales allows us to study the centrality of nodes and the structure of neighborhoods at different level of details. Present results are related to the specific DNS realization performed. However, since the turbulent channel flow analyzed is statistically stationary, we expect similar results from other DNS runs, provided $Re_{\tau}$ is the same.
	
	\subsection{Global scale analysis}\label{subsec:global}
	
\begin{figure*}[ht]
	\centering
	\includegraphics[width=.98\linewidth]{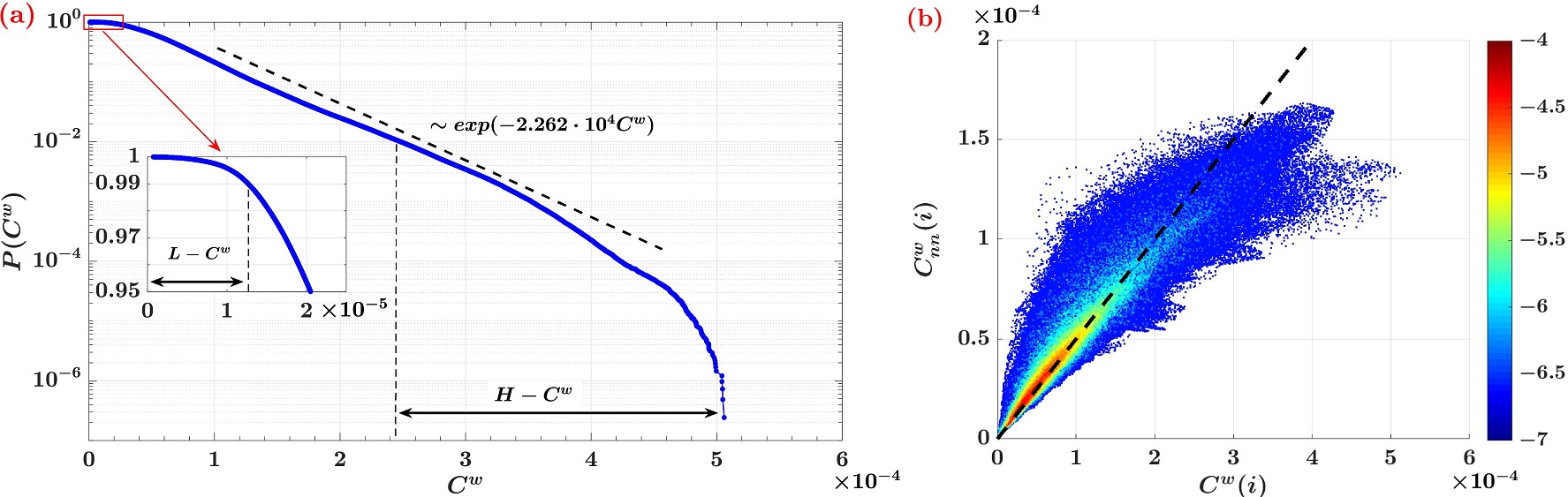}
	
	\caption{Global analysis of the network built on the streamwise velocity component. (a) Cumulative $C^w$ distribution, $P(C^w)$, and exponential fitting. The inset is a zoomed view for small $C^w$ values, indicated by the red box. The ranges of $H-C^w$ and $L-C^w$ (shown in the inset) are highlighted. (b) Weighted average nearest neighbors assortativity measure, $C^w_{nn}(i)$, as a function of $C^w(i)$. Colors indicate the joint probability values (in $log_{10}$ scale) of variables $C^w(i)$ and $C^w_{nn}(i)$. The bisector is also displayed as a black dashed line. \label{fig:PVWC_assort_u}}
\end{figure*}

		The global behaviour of the networks is studied by investigating $C^w$, which represents the fraction of volume kinematically connected to a node (see Eq. (\ref{eq:VWC})). First, we focus on the network of the streamwise velocity component. In order to understand how $C^w$ is distributed in the network, in Fig. \ref{fig:PVWC_assort_u}a we show the $C^w$ cumulative probability, $P(C^w)$. The probability to have higher values of $C^w$ decreases exponentially, suggesting there is a relatively small number of nodes that are strongly connected with respect to the other nodes, thus representing the \textit{hubs} of the network. The $P(C^w)$ distribution can be exploited to classify the centrality of nodes in the network. In the following, for the networks of both velocity components, we will define a node as a $H-C^w$ node (i.e., with a high $C^w$ value) if its $C^w$ satisfies $P(C^w)\leq 10^{-2}$ (corresponding to the 99th percentile). On the contrary, we will refer to nodes with a low $C^w$ value as $L-C^w$ nodes, indicating that their $C^w$ value satisfies $P(C^w)=99\times10^{-2}$ (corresponding to the 1st percentile). $H-C^w$ nodes represent parts of the domain kinematically similar to large portions of the physical domain, in the temporal window considered.

		$C^w$ is a measure of the centrality of nodes in the network, but it is not able to quantify whether the centrality of a node is similar or not to the centrality of its first neighbors. To this aim, a typical metric to investigate the inter-relation among nodes is the assortativity, revealing if nodes tend to link to other nodes with similar or dissimilar $C^w$ values. The average $C^w$ of neighbors of a generic node $i$, $C^w_{nn}(i)$ (see Eq. (\ref{eq:knn_VWC})), as a function of $C^w(i)$ is shown in Fig. \ref{fig:PVWC_assort_u}b. An almost linear relationship holds between $C^w$ and $C^w_{nn}$, displaying that most nodes tend to link to other nodes with quite the same $C^w$ value, resulting in a strongly assortative network. The joint probability between $C^w$ and $C^w_{nn}$ is also evidenced with different colors in Fig. \ref{fig:PVWC_assort_u}b: higher joint probability values concentrate along the bisector and for small $C^w$ values. More in detail, low-$C^w$ nodes tend to have neighbors with similar or higher $C^w$ values, while high $C^w$ nodes tend to link to nodes with similar or slightly lower $C^w$ values. This outcome implies that parts of the domain with (linearly) similar time-series of the streamwise velocity $u$ (i.e., high correlation coefficients values) have also similar neighborhood spatial extensions. In other words, the fraction of volume highly correlated with a node $i$ and the fraction of volume highly correlated with the first neighbors of $i$, are of the same order of magnitude.

		In the network based on the wall-normal velocity time-series, a sharp decay of $P(C^w)$ is found. Therefore, the same definition of $H-C^w$ and $L-C^w$ nodes also holds for the network based on the $v$ component. Moreover, as for $u$, the network based on the wall-normal velocity displays a strong positive assortative behavior. More details can be found in Fig. \ref{fig:PVWC_assort_v}, in Appendix B.
				
		The analysis at global scale points out that hubs are generally rare in the networks and, as all the other nodes, they tend to connect with each others.

	\subsection{Meso-scale analysis}	\label{subsec:meso}

Moving from a global to a meso-scale level of analysis, the structure of the networks as a function of the wall-normal coordinate, $y^+$, is firstly investigated. Due to the symmetrical behaviour of the results with respect to the center of the channel, the plots of the metrics as a function of $y^+$ are shown as averages of both halves of the channel (i.e.,  $y^+\in[0,180]$). Next, the analysis at meso-scale is focused on the most central nodes of the network.

		\subsubsection{Analysis along $y^+$ direction}	
		
\begin{figure*}[ht]
	\centering
	\includegraphics[width=.98\linewidth]{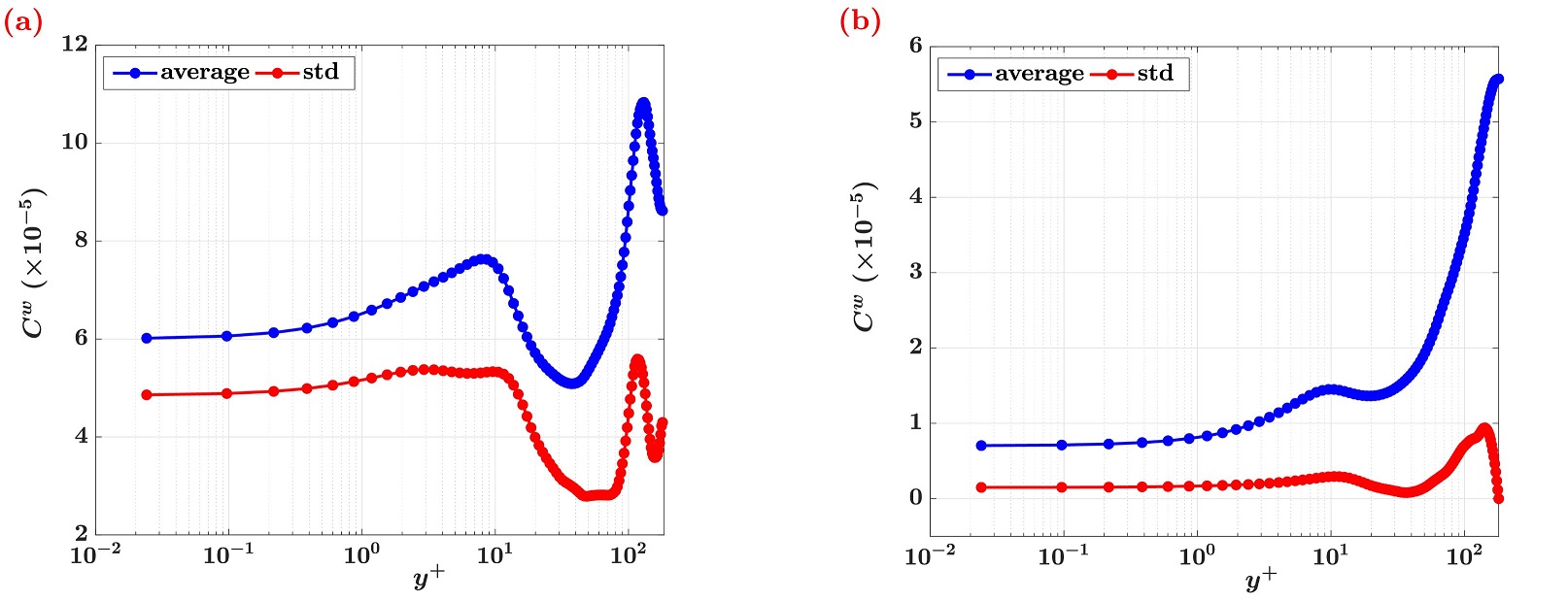}
	
	\caption{Mean and standard deviation values of the volume-weighted connectivity, $C^w$, as a function of $y^+$, and averaged over the two homogeneous directions. (a) Network based on the streamwise velocity; (b) network built on the wall-normal velocity. \label{fig:vwc_y}}
\end{figure*}

		Mean and standard deviation values of $C^w$ in planes at constant $y^+$ are first considered, for both $u$ and $v$. As shown in Fig. \ref{fig:vwc_y}, the local maxima of mean $C^w$ values are found at distances (relatively to each wall) of about $y^+\cong10$ and around the center of the channel. In particular, for the network built on the $u$ component, the highest peak is located at about $y^+\cong120$. Such local peaks of the average values suggest the presence of a large number of $H-C^w$ nodes around those locations. Local maxima of the standard deviation are found at about $3 \lesssim y^+ \lesssim 10$ and $120 \lesssim y^+ \lesssim 180$, that are almost at the same $y^+$ as the local peaks of the average value. This implies that $H-C^w$ nodes increase the variability of the $C^w$ values at these wall-normal locations.

\begin{figure*}[ht]
	\centering
	\includegraphics[width=.98\linewidth]{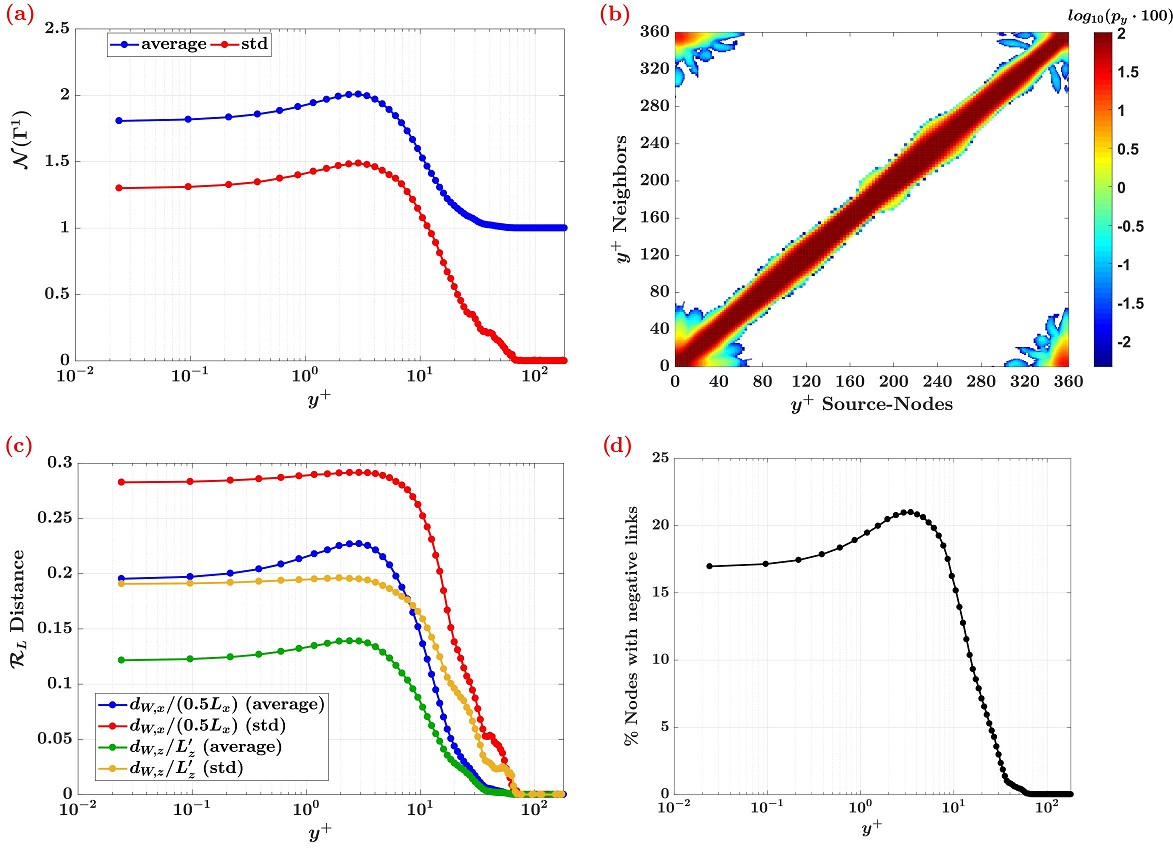}
	\hfill
	\caption{Characterization of the first neighborhood, $\Gamma^1$, as a function of $y^+$ for the network built on streamwise velocity. (a) Mean and standard deviation values of the number of regions, $\mathcal{N}(\Gamma^1)$, as a function of the distance to the wall, $y^+$, and averaged over the two homogeneous directions. (b) The probability (in log scale) that a source-node at a given $y^+$ is linked to a neighbor at another $y^+$ value. (c) Mean and standard deviation of weighted physical distances between nodes at fixed $y^+$ and their long-range neighbors, averaged over the two homogeneous directions. The distances are normalized with the maximum distances, $0.5L_x=2\pi$ and $L_z'=25/36\pi$, in the $(x,z)$ directions. Due to the periodicity of the domain, the maximum distance in the $x$-direction is $L_x/2$ instead of $L_x$. (d) The fraction of nodes at constant $y^+$ with at least one negatively correlated link. \label{fig:y_structure}}
\end{figure*}	

\begin{figure*}[ht]
	\centering
	\includegraphics[width=.98\linewidth]{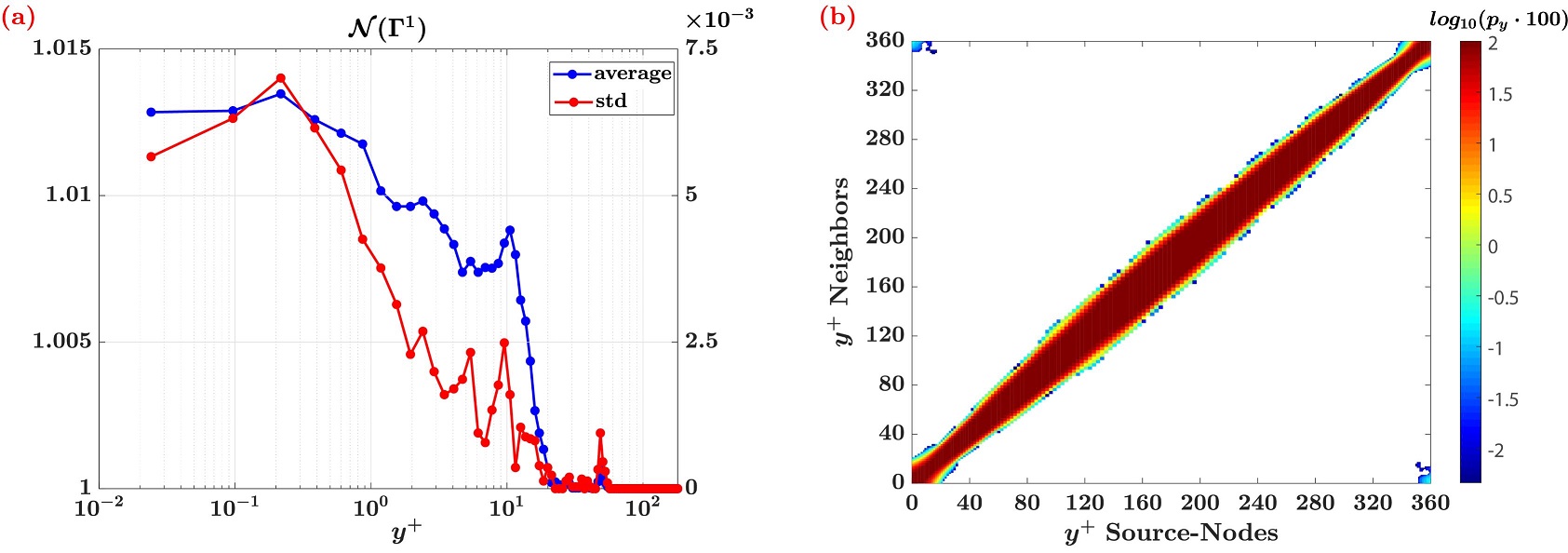}			
	
	\caption{Meso-scale results for the network built on wall-normal velocity. (a) Mean (left axis) and standard deviation (right axis) values of the number of regions, $\mathcal{N}(\Gamma^1)$, as a function of the distance to the wall, $y^+$, and averaged over the two homogeneous directions. (b) The probability (in log scale) that a source-node at a given $y^+$ is linked to a neighbor at another $y^+$ value. \label{fig:V_meso}}
\end{figure*}			
			
		Next, we investigate the relation between nodes and their first neighbors, $\Gamma_i^1$. In particular, we inspect where the first neighbors of a node $i$ are located in the domain, whether they spread all over the domain or there is some kind of spatial organization. To this end, we characterize the first neighborhood, $\Gamma^1$, of nodes at different $y^+$, through its most significant features, such as the number of regions in which the neighborhood is divided, its position and correlation sign with respect to each corresponding node. 
		
		First, the network built on the streamwise velocity, $u$, is considered. As for the neighborhood repartition, we evaluate the average and the standard deviation values of the number of regions, $\mathcal{N}(\Gamma^1)$, formed by $\Gamma^1$ neighbors of nodes at fixed $y^+$. We recall that a region is defined as a set of geometrically connected nodes, where the geometrical connectivity is a six orthogonal connectivity in the Cartesian discretization. As shown in Fig. \ref{fig:y_structure}a, the first neighborhood of nodes close to the wall tends to be composed of more than one region (non-integer values are due to the averaging), while from $y^+\cong70$ up to the center of the channel, the first neighbors $\Gamma^1$ form only one region (with standard deviation equal to zero).

		\noindent To explore the location of the first neighbors in the domain, the probability that an arbitrary \textit{source-node} at a fixed $y^+$ plane has a neighbor at another $y^+$ value is shown in Fig. \ref{fig:y_structure}b. Nodes at any $y^+$ have first neighbors close to themselves (diagonal part of the plot), but only nodes at a distance $y^+\lesssim 70$ from one wall have non-zero probability values also near the other wall. Therefore, in the network built on the $u$ component there are wall-wall links (both between nodes close to the same wall and at different walls) and center-center links, but there are no direct wall-center connections. These long-range regions, $\mathcal{R}_L$, are also present in the homogeneous directions, as shown in Fig. \ref{fig:y_structure}c. Here, the spatial separation in the streamwise and spanwise directions between nodes at fixed $y^+$ and their long-range neighbors, is investigated by evaluating the average physical distances, $\left\langle d_{W,x}\right\rangle$ and $\left\langle d_{W,z}\right\rangle$ (see Eq. (\ref{eq:dist_mean})). Fig. \ref{fig:y_structure}c shows the average and standard deviation values of $\left\langle d_{W,x}\right\rangle$ and $\left\langle d_{W,z}\right\rangle$ for nodes in planes at constant $y^+$: moving from the center towards the wall, the long-range neighbors of a node tend to be located at increased distance (on average) in the streamwise and spanwise directions. Moreover, long-range neighbors of nodes at the same $y^+$ are quite scattered in the $x$ and $z$ directions, as suggested by the high values of the standard deviation in Fig. \ref{fig:y_structure}c.

\noindent From the two-point spatial autocorrelation definition \cite{kim1987turbulence, sillero2014two}, it is straightforward expecting that some of the first neighbors of each node $i$ in the network are always located close to $i$, forming the short-range region, $\mathcal{R}_S$. This can be seen in Fig. \ref{fig:y_structure}b, where the highest probability values are in the diagonal part of the plot, and in Fig. \ref{fig:y_structure}(a,c) for $y^+\gtrsim 70$, where the neighborhood $\Gamma^1$ coincides with $\mathcal{R}_S$. Instead, what is not trivial is the emergence of long-range links in all directions, more specifically inter- and intra-wall links occurring for $y^+\lesssim 70$. By analogy with the climate analyses, we refer to long-range links as \textit{teleconnections} \cite{wallace1981teleconnections, Nigam201590, zhou2015teleconnection, kittel2017global, hlinka2014non, arizmendi2017enso}. In atmospheric sciences, teleconnections indicate climate relations (in terms of temperature, rainfall, pressure or other quantities) between geographically remote regions, farther than the correlation length scale of the variable. Climate teleconnections are mainly caused by the energy transport and propagation of waves, providing information about the recurrence of climate variability of distant locations. Here, the emergence of teleconnections of the streamwise velocity can be interpreted as the footprints of the top-down interactions, which similarly act from the outer layer to both near-wall regions \cite{Hwang2016,Hutchins2007}. In fact, teleconnections are always individuated between regions close to the two walls (or close to the same wall), revealing an analogous response of the two wall regions to the large-scale structures (i.e., turbulent structures with size of the order of the integral space scale). On the contrary, teleconnections are never found between inner and outer layer regions, where the interplay dynamics are deeply different one from each other. Therefore, complex networks are able to unveil the presence of teleconnections, which are usually hidden by the spatial averaging of the correlation coefficient values. Teleconnections create a texture of links (highlighted by the network metrics) between distant locations, in which similar (streamwise) kinematic information persists in time. This result is the main difference with respect to other approaches in the turbulence research, where the usual spatially-averaged correlation only retains average information about the spatial behavior of the correlation field.
			
			To complete the analysis as a function of $y^+$, we examine the distribution of the sign of the correlation coefficient of links between nodes and their first neighbors. By construction, links in the network are active if the absolute value of the correlation coefficient, $C_{i,j}$, is above $\tau=0.85$, but links can have either negative or positive $C_{i,j}$ values. In Fig. \ref{fig:y_structure}d, the fraction of nodes at fixed $y^+$ with at least a negative-correlated neighbor is shown as a function of the wall-normal coordinate. Negative-correlation links are found (in the network based on $u$) only for $y^+\lesssim70$, with a peak at $y^+=3.5$ that coincides with the peak of the average number of regions in Fig. \ref{fig:y_structure}a. In particular, among the nodes with negatively correlated neighbors, the occurrence of negative links is (on average) about $10\%$ of total links. Based on what observed so far, we can infer that negative correlation links are possible due to the presence of teleconnections (i.e., $y^+\lesssim70$), while short-range links are only activated by positive correlation values (as for $y^+\gtrsim70$).

		For the network built on the wall-normal velocity component, the number of regions, $\mathcal{N}(\Gamma^1)$, of the first neighborhood is shown in Fig. \ref{fig:V_meso}a, while the probability that an arbitrary source-node has a neighbor at another $y^+$ value is illustrated in Fig. \ref{fig:V_meso}b. In analogy with Fig. \ref{fig:y_structure}a, values of the average number of regions greater than one are found only close to the wall. However, in this case, the average values are close to one, with very low standard deviation. This implies the substantial absence of long-range links in the network of the wall-normal component, i.e., teleconnections rarely appear. This behavior is also confirmed by the probability to have a neighbor at a given $y^+$. As shown in Fig. \ref{fig:V_meso}b, most of the nodes connect with nodes close to them, and only few points very close to the wall have teleconnected neighbors close to the other wall.

		\subsubsection{Analysis of the most central nodes}

		We here focus on the hubs of the networks (i.e., $H-C^w$ nodes) and their first neighbors, to understand whether they form spatial patterns and how the neighborhood of such hubs is structured. As in the previous section, the network built on the streamwise velocity, $u$, is explored first.

\begin{figure*}[ht]
	\centering
	\includegraphics[width=.98\linewidth]{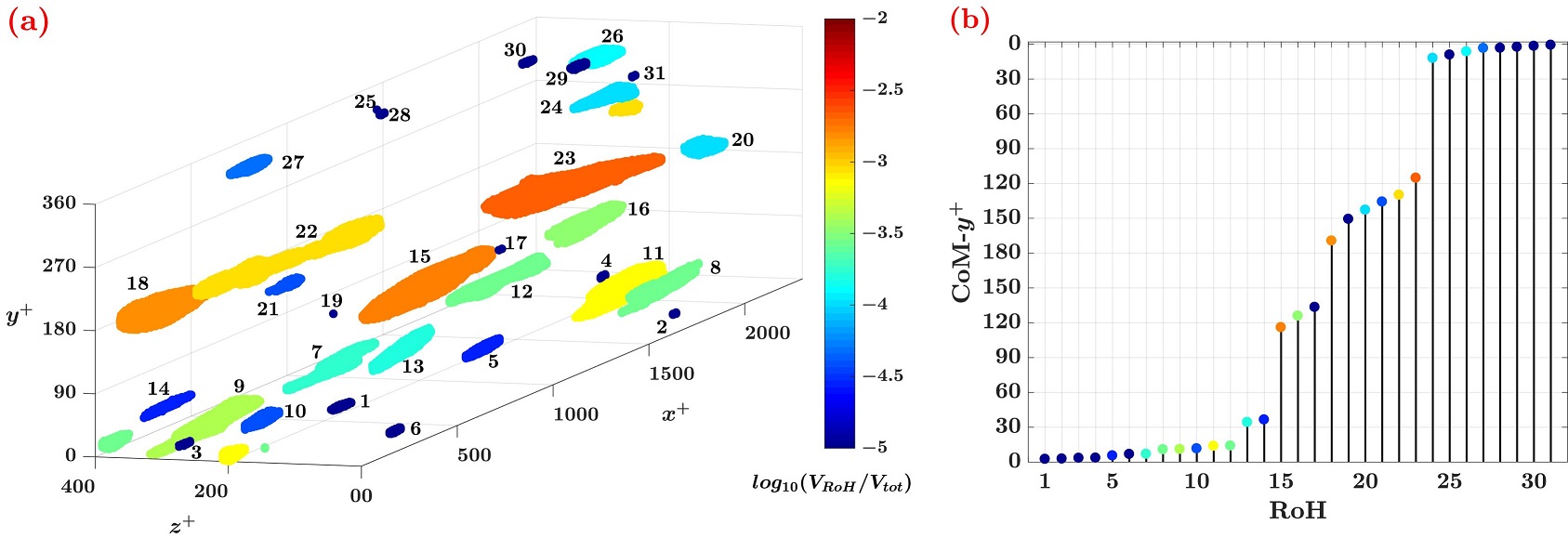}
	\caption{(a) 3D view of $H-C^w$ nodes, namely regions of hubs, RoHs, for the network built on streamwise velocity. Color scale refers to the fraction of volume occupied by distinct RoHs, $V_{RoH}/V_{tot}$. RoHs are labeled for increasing values of the wall-normal coordinate of their center of mass. Periodicity of the domain in the $x$ direction is visible from the RoHs labeled $8$,$11$,$12$ and $22$. (b) Weighted $y^+$-component of the center of mass, CoM-$y^+$, of the RoHs shown in panel (a), where CoM-$y^+=\sum_i{\left(y_i^+ V_i/V_{tot}\right)}/\mathcal{N}(RoH)$, with $i\in$RoH and $\mathcal{N}(RoH)$ the number of nodes in each RoH. Colors indicate the fraction of volume of each RoH as shown in (a). The reader is referred to the online version for a high color resolution. \label{fig:H_VWC_y_3D}}
\end{figure*}

		The spatial location of the $H-C^w$ nodes is shown in a 3D view in Fig. \ref{fig:H_VWC_y_3D}a. Highly connected nodes are not scattered in the domain, but they tend to locally group into clusters elongated in the streamwise direction (the longest one with a streamwise extension of about $\Delta x^+\simeq 600$). According to the definition of region, $\mathcal{R}$, the $H-C^w$ nodes form in this case 31 regions, which we call regions of hubs (RoHs). It is important to remind that nodes in the same RoH are not necessarily all linked to each other; some of them may be linked, but the RoHs merely identify groups of high $C^w$ nodes belonging to the same spatially-connected region. Such RoHs have different sizes, as illustrated in Fig. \ref{fig:H_VWC_y_3D}a where colors indicate the fraction of volume occupied by each RoH, namely $V_{RoH}/V_{tot}$. The RoHs are present at different $y^+$, as displayed in Fig. \ref{fig:H_VWC_y_3D}b, in which the wall-normal coordinate of the center of mass of each RoH is shown. From here it emerges that the presence of the biggest RoHs (around $y^+\simeq15$, RoHs 7-12, 24, 26, and $y^+\simeq120$, RoHs 15, 16, 22, 23) is the main responsible of the local peak values of $C^w$, previously observed in Fig. \ref{fig:vwc_y}a.

	\noindent The occurrence of similar patterns of RoHs throughout the domain is a remarkable outcome. In fact, one would expect different spatial patterns of $H-C^w$ nodes at different $y^+$, because the two-point correlation of the streamwise velocity changes along $y^+$ (see the average behaviour at different $y^+$ in Fig. \ref{fig:Umean_autocorr}b). Instead, although the network is based on the two-point correlation, it is able to advance the level of information by retaining, all at once, the multi-point effects of active links in all directions. This outcome emphasizes the potential of the complex network approach to enrich the spatial characterization of wall turbulence.
		
		$H-C^w$ nodes of the network built on the $v$ component also tend to form RoHs elongated in the streamwise direction, but they appear around the center of the channel, as already shown in Fig. \ref{fig:vwc_y}b (more details are reported in Fig. \ref{fig:apdx_RoH_V}, in Appendix B). Therefore, the elongated shape of the RoHs is not strongly dependent on the variable selected, but it can be seen as an effect of the mean flow in the streamwise direction. Turbulent structures are indeed advected downstream by the mean flow in the $x$ direction, and the typical timescale in which turbulence evolves is larger than the advection timescale (this is the so-called Taylor's hypothesis \cite{taylor1938production, smits2011high}). Consequently, this feature is not detected for high-degree nodes in the correlation network for homogeneous isotropic turbulence \cite{scarsoglio2016complex}.

\begin{figure*}[ht]
	\centering
	\includegraphics[width=.98\linewidth]{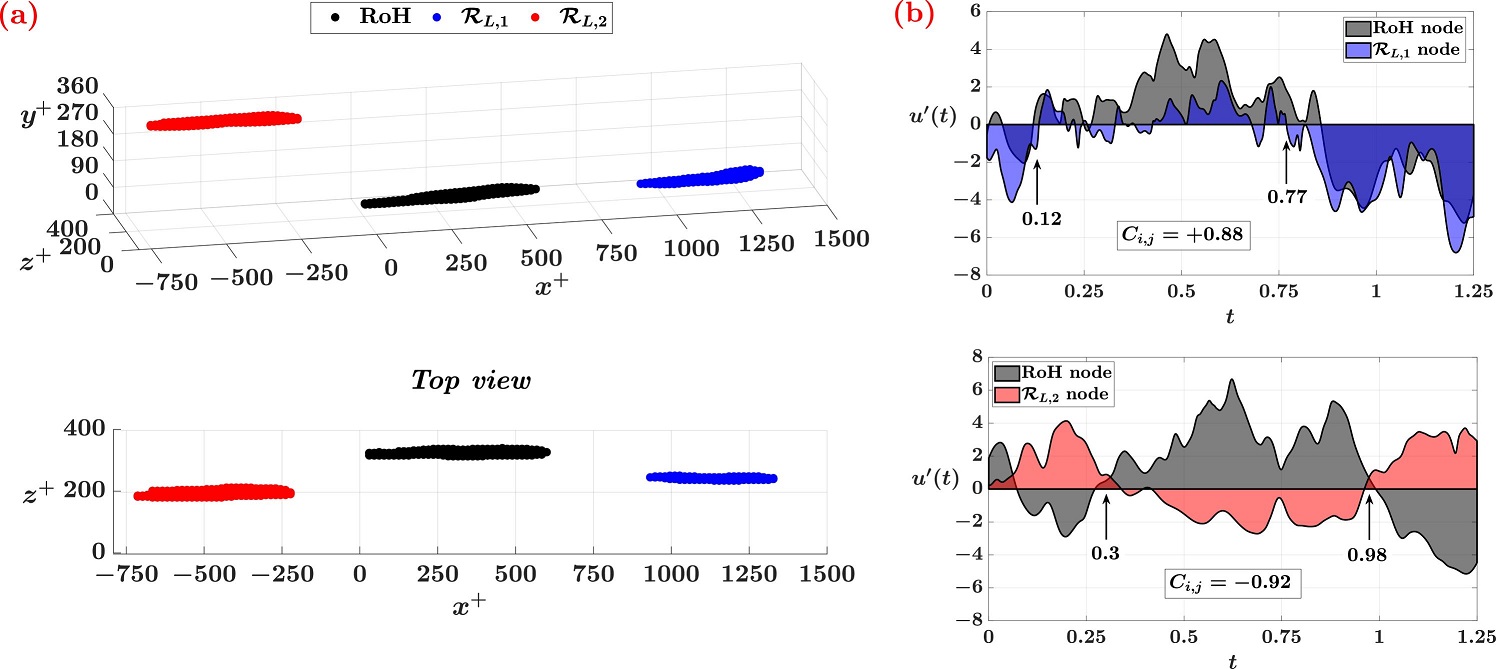}			
	\caption{(a) 3D view of nodes in an RoH (black) and two $\mathcal{R}_{L}$ regions formed by long-range neighbors of nodes in the RoH (only a fraction of nodes in the RoH is actually linked to each $\mathcal{R}_{L,1-2}$ region, because these regions are the union, and not the intersection, of the neighborhoods of the nodes in the RoH). The $y^+$ values of the center of mass of each $\mathcal{R}_{L}$ are $y^+\simeq 39$ and $y^+\simeq 353$ (i.e., $y^+\simeq 7$ to the closer wall), while the $y^+$ of the RoH is $y^+ \simeq 11$. The blue and red colors indicate a positive and negative correlation with the RoH, respectively. (b) Time-series of the streamwise velocity fluctuations, $u'$, of two different pair of nodes; times are in terms of $H/u_\tau$. The black shaded series correspond to nodes in the RoH, while blue and red shaded series correspond to nodes in $\mathcal{R}_{L,1}$ and $\mathcal{R}_{L,2}$, respectively. The values of the correlation coefficients and the temporal limits of the maximum time-interval with the same $u'$ sign for the two pairs of nodes, are also reported. \label{fig:streaks} }
\end{figure*}

		Regarding the neighborhood of the most central nodes, since the hubs are clustered into RoHs, we consider the first neighborhood of all nodes in the RoHs. In particular, here we focus on the $u$-based network, since non-trivial teleconnection patterns represent the most notable outcome and they are mainly found in this network. By exploring the long-range neighborhoods of nodes in the RoHs, we found that they exhibit a peculiar behavior. Not only first neighbors of nodes at $y^+\lesssim70$ form long-range regions (as already observed in Fig. \ref{fig:y_structure}), but here we find that hubs belonging to the same RoH generate long-range regions which are physically close one to each other. In other words, long-range neighbors of nodes in the same RoH are not scattered in the domain but constitute themselves spatially-connected regions. In Fig. \ref{fig:streaks}a we show an example of RoH (depicted in black and corresponding to the ninth RoH in Fig. \ref{fig:H_VWC_y_3D}) and two regions ($\mathcal{R}_{L,1}$ and $\mathcal{R}_{L,2}$, depicted in blue and red, respectively) formed by the union of long-range neighbors of the nodes in the RoH. As can be seen, the regions $\mathcal{R}_{L,1-2}$ inherit the same elongated shape in the streamwise direction, $x$, and similar volumes of the corresponding RoH. Such a behavior is found for all RoHs and their long-range neighbors; another example can be found in the Supplemental Material \cite{supmat}.
		
		\noindent This outcome extends the meaning of teleconnections from nodes to regions: in the domain, there are regions of highly connected nodes (i.e., the RoHs) that are strongly linked with distant spatially-connected regions. These pairs of \textit{teleconnected regions}, therefore, represent near-wall portions of the domain tightly correlated over time from the streamwise velocity point of view, i.e. spatially extended regions sharing similar dynamics in time. A movie of a representative RoH and the corresponding teleconnected regions is reported in the Supplemental Material, see Movie SM1.

	\noindent A further element to characterize the neighborhood of the hubs (or RoHs) is the sign of the correlation of links. For a generic node $i$, we find that the first neighbors belonging to a region are either all positively or all negatively correlated with $i$. In other words, for any node $i$ in the network, the regions formed by its neighbors are never partially positively/negatively correlated with $i$, but always exhibit the same correlation sign. This means that, considering again long-range links only, each region formed by the union of long-range neighbors of nodes of an RoH, has a unique correlation sign with the corresponding RoH. For example, in Fig. \ref{fig:streaks}a, the nodes of the selected RoH (colored in black) are all positively correlated with the corresponding neighbors in $\mathcal{R}_{L,1}$ (colored in blue), and all negatively correlated with those in $\mathcal{R}_{L,2}$ (colored in red).
		
	The meso-scale analysis evidences the presence of spatially-connected regions of highly linked nodes (RoHs), both in the near-wall and outer layer. However, only near-wall nodes are characterized by teleconnections. In particular, pairs of teleconnected regions (not only pairs of nodes) are found, which correspond to regions of fluid moving with similar streamwise velocity in time, characterized by either positive or negative correlations. A possible physical explanation of the teleconnection patterns emerges from the inspection of the streamwise velocity time-series. To this aim, we arbitrarily select two pairs of neighbors from the regions shown in Fig. \ref{fig:streaks}a: the first pair composed of a node in RoH and a neighbor in $\mathcal{R}_{L,1}$, and the second pair composed of a node in RoH and a neighbor in $\mathcal{R}_{L,2}$. For both pairs, in Fig. \ref{fig:streaks}b the time-series of the streamwise velocity fluctuations, $u'=u-U$, are plotted. It can be noted that the streamwise fluctuations have mostly the same sign for large time-intervals: for the pair involving the node in $\mathcal{R}_{L,1}$, during the time interval $\Delta t\approx |0.77-0.12|=0.65 H/u_\tau$, while for the pair involving the node in $\mathcal{R}_{L,2}$, during the time interval $\Delta t_i\approx |0.98-0.3|=0.68 H/u_\tau$. This behaviour is typical of high/low speed coherent streaks, that is alternating near-wall regions of positive/negative velocity fluctuations, with an average spanwise separation of $\Delta z^+\sim 100$ and streamwise lengths $\Delta x^+\sim 10^3\div10^4$ \cite{jimenez2013near}. By applying the Taylor hypothesis of frozen turbulence \cite{taylor1938production} and the typical near-wall convective velocity, $U_c^+\approx10$ \cite{geng2015taylor}, we estimate that the range, $\Delta x^+$, corresponding to the time intervals $\Delta t\approx 0.65$, is $\Delta x^+=(\Delta t  Re_\tau)U_c^+\approx 1200$, which is in agreement with the typical streamwise elongation of the streaks. The presence of near-wall teleconnections with both positive and negative correlation sign can be thus interpreted as an imprint of turbulent coherent structures with time-scale of the order of the temporal window considered. In particular, the complex network approach is able to provide a high level of spatial information (i.e., spatial position, shape and size) of such coherent patterns, thus enriching the spatial characterization of wall-bounded turbulent flows.
	
	\subsection{Local scale analysis}\label{subsec:local}
		
		The local scale analysis is useful to focus on specific nodes and highlight how the kinematic information spreads through the domain starting from these nodes. Given a source-node, we inquire what is the \textit{correlation path} in the network linking that node to all the others. Specifically, in this Section we only consider the nodes of the network based on the $u$ component, since only this network exhibits non-trivial teleconnection patterns. In fact, the network built on the correlation coefficient of the streamwise velocity time-series is not just a trivial collection of correlated points, but it represents a pattern of (linear) inter-connections among nodes. Therefore, the network can be interpreted as a structure of links over which the kinematic information moves throughout the domain. In particular, if a node $i$ is linked to a node $j$ (i.e., by hypothesis, they are strongly correlated) and the node $j$ is linked to a third node $k$ (but $i$ is not linked to $k$, i.e. $A_{i,k}=0$), then the kinematic information indirectly flows from $i$ to $k$ by means of $j$ (i.e., $i$ and $k$ are indirectly linked). To this end, we analyze the cumulative neighborhoods, $\Gamma^{N,c}$, and the shortest paths of nodes representing extreme features, that is nodes with high/low $C^w$ values and close/far from the wall. Three pairs of nodes ($c$, $w1$, $w2$) are selected, where each pair contains one $L-C^w$ node and one $H-C^w$ node: $c$, is taken at the center of the channel at $y^+=180$; $w1$ and $w2$ are taken close to each of the two walls, at $y^+\simeq 3.5$. The behaviour of all other nodes with intermediate $y^+$ and $C^w$ values lies in between.

\subsubsection{Analysis of the cumulative neighborhoods}

\begin{figure*}[ht]
	\centering
	
	\includegraphics[width=.98\linewidth]{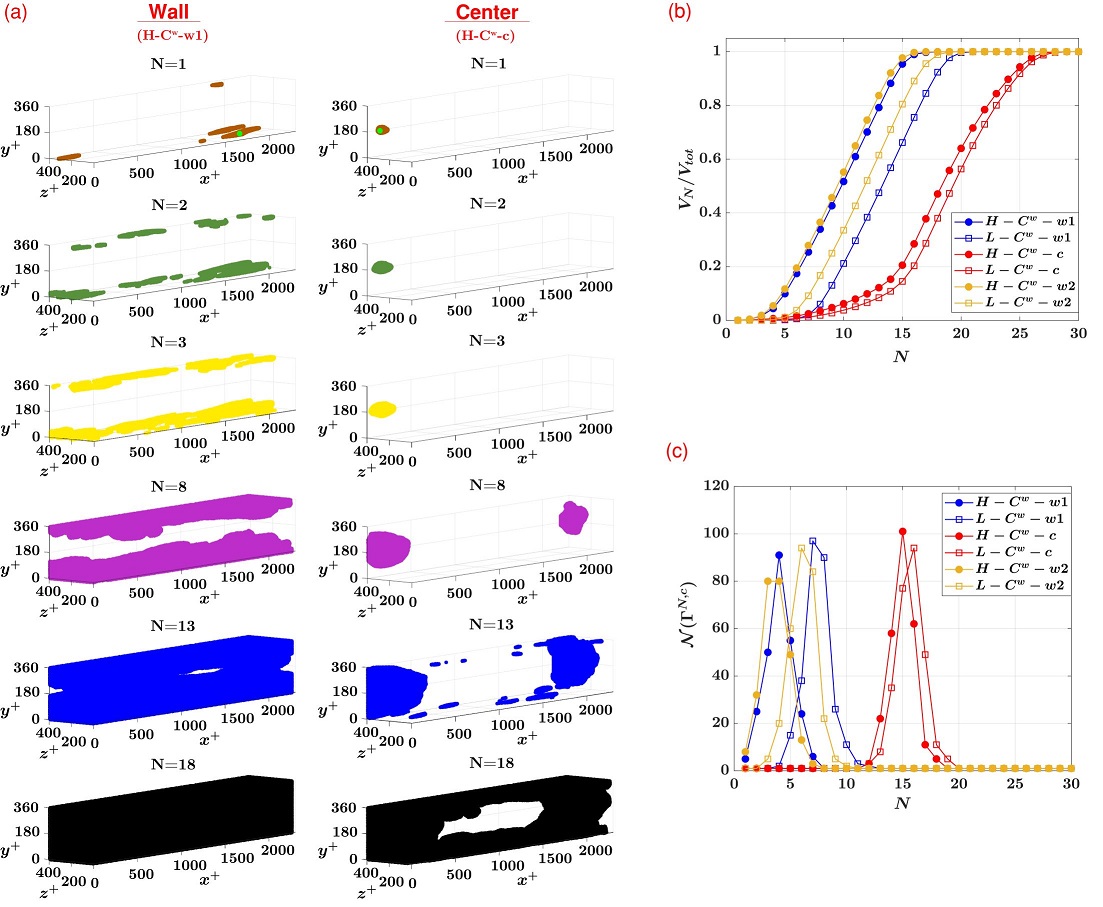}			
	\caption{(a) 3D views of the $N$-th cumulative neighborhoods of $H-C^w$ nodes (marked in green in the $N=1$ panels). (left) node close to the wall at $y^+\simeq 3.5$; (right) node at the center of the channel, $y^+=180$. (b)-(c) Fraction of volume, $V_N/V_{tot}$, and number of regions, $\mathcal{N}(\Gamma^{N,c})$, of the $N$-th cumulative neighborhoods for the three selected pairs of nodes ($w1$, $w2$, $c$). \label{fig:cum_neigs}}
\end{figure*}

	\noindent The behaviour of the successive neighborhoods of a source-node is ruled by several factors, such as the number and the size of the regions of the first neighborhoods, and the distance of the neighbors from the source-node. To visualize the differences between source-nodes close to the walls and at the center of the channel, in Fig. \ref{fig:cum_neigs}a we plot the positions of the $N$ cumulative neighborhoods, $\Gamma^{N,c}$, with $N=\left\lbrace 1,2,3,8,13,18\right\rbrace$, for the nodes $H-C^w-w1$ and $H-C^w-c$. The resulting spatial expansion of $\Gamma^{N,c}$ for nodes at different $y^+$ appears completely different. It is worth noting that the network is built on a periodic computational domain in the $x$ direction, which markedly affects the three-dimensional views of the $\Gamma^{N,c}$ (e.g., see the $N=8$ right panel in Fig. \ref{fig:cum_neigs}a). By focusing firstly on the node close to the wall (left panels in Fig. \ref{fig:cum_neigs}a), the first three neighborhoods are all close to both walls, revealing the occurrence of teleconnections which are scattered in the $(x,z)$ directions. In particular, the pattern of these neighborhoods is very anisotropic, since it is quite elongated in the streamwise direction. Neighborhoods for $N>3$ also include nodes close to the walls first (see for example the panel at $N=8$), starting to incorporate nodes at higher $y^+$ values and finally approaching the center of the channel. Therefore, in this case, the cumulative neighborhoods move progressively from the walls to the center of the channel, implying a very high connectivity among nodes close to the walls, where teleconnections play a crucial role. On the other hand, for the node $H-C^w-c$ (right panels in Fig. \ref{fig:cum_neigs}a), the first neighborhoods are all close to it, they are almost isotropic in the $(y,z)$ directions and a bit elongated along the $x$ direction. Only from the $N=13$ neighborhood, the nodes close to the walls start to be included. From this step onward, all the nodes close to the walls are first included in the successive neighborhood expansion (see right panel in Fig. \ref{fig:cum_neigs}a for $N=18$), and only later all the other nodes at higher $y^+$ are covered. Therefore, the behaviour displayed in Fig. \ref{fig:cum_neigs}a indicates that, either by considering a node close to the wall or at the center of the channel, the $\Gamma^{N,c}$ expansion in the $y^+$ direction does not occupy the volume of the domain in a monotonic way. In fact, the near-wall regions tend to be included faster, while the central part of the domain is covered afterwards.

	Panels (b) and (c) of Fig. \ref{fig:cum_neigs} further quantify the structure of successive neighborhoods for the three pairs of nodes ($c$, $w1$, $w2$), through the fraction of volume, $V_N/V_{tot}$, and the number of regions, $\mathcal{N}(\Gamma^{N,c})$, occupied by the first $N$ neighborhoods as a function of $N$, respectively. By starting with the fraction of volume (panel b), for both pairs of nodes $w1$ and $w2$ (see blue and orange curves, respectively) the expansion of the neighborhoods is initially much faster than the expansion for the pair $c$ (red curves). However, at intermediate $N$ values (i.e., from around $N=5$ for the $w1$ and $w2$ pairs, and $N=15$ for the $c$ pair), the $\Gamma^{N,c}$ of all source-nodes tend to growth almost linearly with similar slopes: these ranges of $N$ values correspond to a wall-normal stratified increase of $V_N/V_{tot}$. From this range on, the $N$-th neighborhoods are composed of almost planar layers of nodes parallel to the wall (e.g., Fig. \ref{fig:cum_neigs}a, left panels at $N=8,13$ and right panel at $N=18$). As for the number of regions (panel c), the peaks of $\mathcal{N}(\Gamma^{N,c})$ occur at low $N$ for the $w1$ and $w2$ pairs, while for the $c$ pair the maximum values of $\mathcal{N}(\Gamma^{N,c})$ are attained at higher $N$. Values of $\mathcal{N}(\Gamma^{N,c})$ greater than one are localized in specific ranges of $N$ associated to the inclusion of nodes close to the walls (with consequent appearance of teleconnections), while for the remaining $N$ values $\mathcal{N}(\Gamma^{N,c})=1$. For both the fraction of volume and the number of regions, pairs of nodes $w1$ and $w2$ show the same overall behaviour as a function of $N$, which is faster than the one observed for the $c$ pair. Within a pair of nodes, the behaviour of the $L-C^w$ node is similar but slower than the corresponding $H-C^w$. This is in line with the assortativity plot in Fig. \ref{fig:PVWC_assort_u}b: nodes with low $C^w$ values are more likely to be linked to nodes with similar $C^w$ and some more steps are required to reach $H-C^w$ nodes, which are connected to a larger fraction of the domain.
	
		The analysis of the $\Gamma^{N,c}$ neighborhoods provides insights into the kinematic information flow, evidencing that: (i) nodes in the near-wall regions (indicatively, $y^+\lesssim70$) are strongly inter-connected for low $N$ values, creating anisotropic textures of teleconnections, that result in a very effective kinematic information spreading; (ii) nodes around the center of the channel (indicatively, $y^+\gtrsim70$) display localized high-correlation patterns for low $N$ values, similar to those extracted in homogeneous isotropic turbulence \cite{scarsoglio2016complex}; (iii) high $C^w$ nodes are the most central in the network, not only relative to the first neighborhood but also in relation with the whole network.
	
		\subsubsection{Analysis of the shortest paths}

		To conclude the analysis at the local-scale level, we explore the shortest paths between nodes at different wall-normal locations. We recall that a shortest path is the path of minimum cost between two nodes, where the cost represents the \textit{shortest path distance}. If the links in the network are weighted (i.e., a scalar value is assigned to each link), the shortest path distance corresponds to the minimum value of the sum of the link-weights between two nodes. An appropriate metric for weighting links and evaluating the shortest path is the distance $\mathcal{D}_{i,j}=\sqrt{2(1-|C_{i,j}|)}$, which fulfills the three axioms defining a metric \cite{mantegna1999hierarchical} and highlights the paths with high (in modulus) correlation values, $C_{i,j}$. Since each time-series of length $T$ can be viewed as a vector, $\textrm{x}$, in a Euclidean $T$-dimensional space, the distance between two series, $\textrm{x}_i$ and $\textrm{x}_j$ (normalized with the local mean and standard deviation), is $\mathcal{D}_{i,j}\sim\sqrt{\sigma_{\bm{\textrm{x}_i}}^2+\sigma_{\bm{\textrm{x}_j}}^2-2C_{i,j}}$, where $\sigma_{\bm{\textrm{x}}}^2=1$ is the variance of the two normalized series. Moreover, since the direction of the links is not taken into account in this work, the shortest path starting from node $i$ and arriving to node $j$, is the same as starting from $j$ and arriving to $i$ (i.e.,  the order of the end-nodes of the shortest path is not relevant).

\begin{figure*}[ht]
	\centering
	\includegraphics[width=\linewidth]{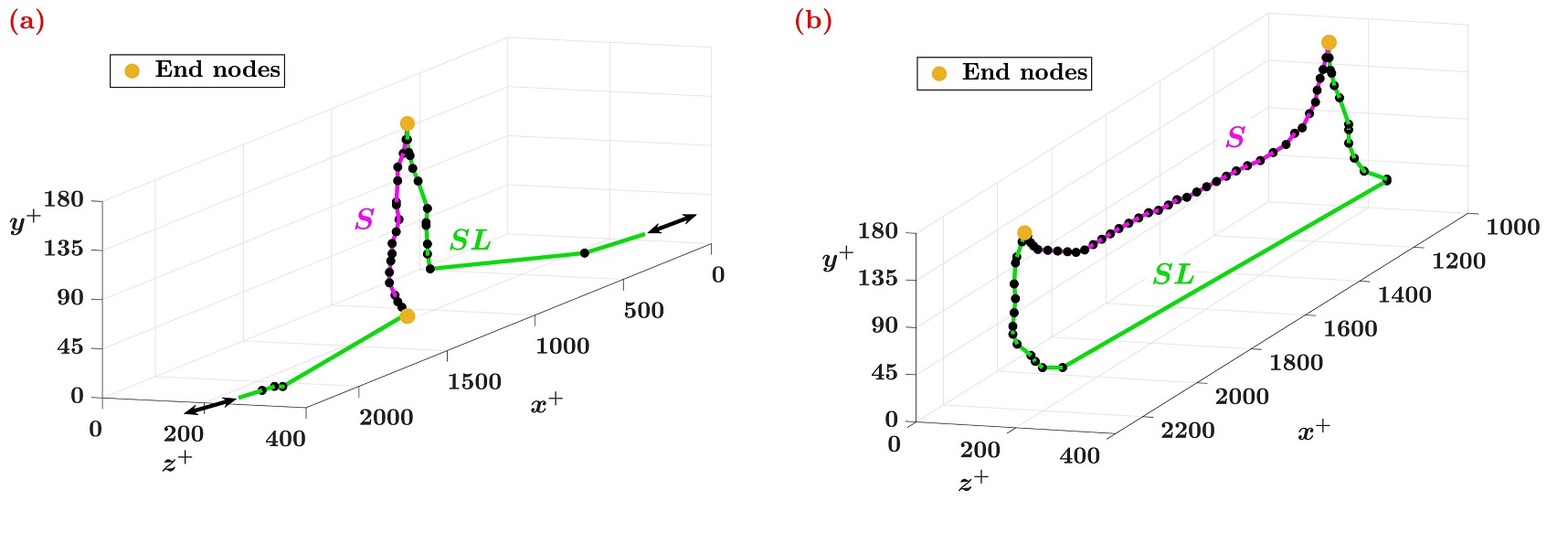}
	
	\caption{Shortest paths in the two configurations: $SL$, i.e., short/long-range links (colored in green); $S$, i.e. only short-range links (colored in magenta). The end-nodes of the paths are depicted in orange, while intermediate nodes are reported in black. (a) End-nodes are at different $y^+$, one close to the wall, at coordinates $(x^+,y^+,z^+)=(1139,3.5,197)$, and the other at the center of the channel, i.e., $(x^+,y^+,z^+)=(1139,180,197)$. The black arrows indicate the periodicity of the domain in the $x$-direction. (b) Both end-nodes are at the center of the channel, at coordinates $(x^+,y^+,z^+)=(1139,180,197)$ and $(x^+,y^+,z^+)=(2262,180,197)$. The reader is referred to the online version for a high color resolution. \label{fig:shortest_p}}
\end{figure*}
	
\begin{figure*}[ht]
	\centering
	\includegraphics[width=\linewidth]{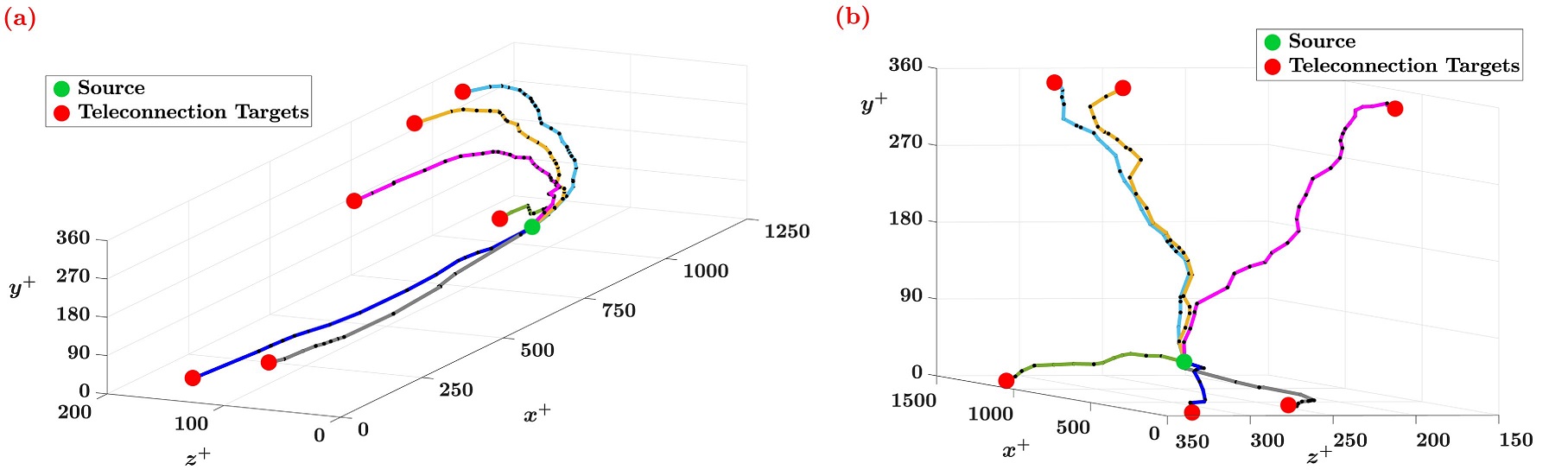}	

	\caption{Example of shortest paths between a source-node close to the wall and six teleconnected neighbors of it, shown through two 3D views. Different colors refer to different targets, while black points indicate intermediate nodes in the paths. The reader is referred to the online version for a high color resolution. \label{fig:telecon_paths}}
\end{figure*}
	
		We analyze two configurations of shortest path, by selecting either short/long-range links or short-range links only. In the first configuration, indicated as $SL$, both short- and long-range links are considered (i.e., the network as it was built) and nodes at different $y^+$ are investigated. Specifically, two pairs of nodes are selected: (i) an end-node of the path close to the wall and the other end-node at the center (see Fig. \ref{fig:shortest_p}a), and (ii) both the end-nodes at the center of the channel (see Fig. \ref{fig:shortest_p}b). The shortest paths for the $SL$ configuration are shown in green in Fig. \ref{fig:shortest_p}. In order to highlight the effects of the teleconnections in the shortest paths, a second configuration (indicated as $S$) is analyzed, in which only short-range links are considered (i.e., long-range links are removed). The resulting shortest paths are shown in magenta in Fig. \ref{fig:shortest_p}, for the same pairs of end-nodes as the $SL$ configuration. For the $SL$ and $S$ shortest paths shown in Fig. \ref{fig:shortest_p}a, the total cost (equal to $\sum \mathcal{D}_{i,j}$) is $7.55$ and $8.69$, while the number of links is 17 and 18, respectively. For the $SL$ and $S$ shortest paths reported in Fig. \ref{fig:shortest_p}b, the total cost is $12.18$ and $18.57$, while the number of links is 26 and 38, respectively. The difference between $S$ and $SL$ configurations is evident: by including the teleconnections (i.e., the $SL$ configuration) the shortest paths are more complex and involve nodes in the near wall region, as a consequence of the strong connectivity of this part of the domain. Furthermore, the $S$ shortest paths are made up of more links and have a higher total cost than the $SL$ shortest paths. It is remarkable to note that in Fig. \ref{fig:shortest_p}b the $SL$ shortest path resembles the pattern shown in the right panels of Fig. \ref{fig:cum_neigs}a, reaching the wall region before connecting again to the channel center. These aspects emphasize the role of teleconnections as intermediary links for kinematic information flow over long distances, even for pairs of end-nodes at the center of the channel.

		To further underline the importance of teleconnections in the overall topology of the network, we show a particular case of the $S$ configuration, in which the end-nodes of the shortest path are linked with teleconnections. Specifically, the shortest paths are evaluated by selecting as end-nodes a source-node in the near-wall domain (since teleconnections are present for $y^+\lesssim70$) and six different teleconnected neighbors of it. These paths are shown, with different colors, in two 3D views in Fig. \ref{fig:telecon_paths}: they represent the shortest paths connecting the source-node to its long-range neighbors (and vice-versa) if teleconnections were not present. The removal of the long-range links implies that the kinematic information has to flow through several short-range links, demonstrating that the presence of teleconnections enhances the spreading of information in the domain.
	
		The local scale analysis provides a detailed description of the topological and kinematic relations between different physical locations in the domain. In particular, the investigation of $\Gamma^{N,c}$ and the shortest paths reveals the full potential of the network in capturing the spatial information related to the patterns of indirect interactions and teleconnections, which is possible only thanks to a network approach.

 \section{Conclusions}	\label{Sec:conclusions}

		In the present work, the complex network analysis was exploited for the study of a fully-developed turbulent channel flow. A spatial network was built, where nodes represent fractions of volume of the physical domain. The correlation coefficient based on the streamwise and wall-normal velocity components was used to activate links, where only correlation values (in modulus) above a given threshold were considered, thus highlighting the strongest kinematic linear inter-relations. The network structure was analyzed at three levels, namely global scale (i.e., considering all nodes, without any distinction), meso-scale (i.e., dealing with groups of nodes), and local scale (i.e., focusing on single nodes).

		First, the presence of hubs in the networks turned out (that is nodes highly connected to other parts of the domain) and a strong assortative behaviours emerged. The analysis of the network at fixed $y^+$ planes revealed that most hubs tend to be localized (on average) at specific $y^+$ values, both close to the walls and around the center of the channel. Moreover, the first neighbors of nodes at $y^+\lesssim70$ tend to cluster into many spatially-connected regions, while for $y^+\gtrsim70$ first neighbors form only one region. This outcome is much more evident in the network built on the $u$ component than in the network based on $v$. By investigating the spatial separation between nodes at fixed $y^+$ and their neighbors, in the network based on $u$ we observed a recurrence of inter-wall and intra-wall long-range links in all directions, which create a kinematic texture of non-trivial connections. We referred to these long-range links as teleconnections. Considering highly connected nodes in more detail, we found that hubs tend to cluster into $x$-elongated regions, RoHs, for both the networks analyzed. However, only in the network built on the streamwise velocity the RoHs appear both close to the walls and in the channel center, while in the network built on the wall-normal velocity they are confined around the channel center.
				
		 Moreover, the teleconnected neighbors of nodes in the same RoHs (i.e., $H-C^w$ nodes at $y^+\lesssim70$), tend to group into spatially-connected regions (similar to the corresponding RoHs). Therefore, RoHs and the corresponding teleconnected regions constitute strongly correlated near-wall parts of the domain, that turned out to be related to the persistence of streamwise velocity streaks, namely near-wall coherent structures. Finally, to highlight the different ways of kinematic information flow in the domain, we investigated the behaviour of the successive neighborhoods of source-nodes with extremely different features, in the network based on $u$. Source-nodes in the near-wall regions are strongly inter-connected, not only relative to the first neighborhood but even in relation to the whole network, resulting in a very effective kinematic information spreading. This is also highlighted by some shortest paths between pairs of nodes, based on the correlation value of each link.
		
		The proposed network-based approach provides a versatile and powerful framework to study complex systems as turbulent flows, especially in the presence of inhomogeneities. Through the application of the network formalism, a different perspective on wall-bounded turbulent flows is introduced, in which the spatial information is preserved and enriched by the multi-point effects of active links in all directions. In fact, teleconnections between distant near-wall regions have been localized and associated with the temporal persistence of coherent patterns in a straightforward way; this operation may turn out to be a complicated task when other techniques are adopted. By taking advantage of the increasing computation capabilities and based on the present findings, the proposed approach can pave the way for a systematic network-based investigation of the turbulence dynamics. Future research will therefore be focused on Reynolds number effects as well as the physical interpretation of network hubs and teleconnections in wall-bounded turbulent flows.

\begin{acknowledgments}
	This work was sponsored by NWO Exacte en Natuurwetenschappen (Physical Sciences) for the use of supercomputer facilities, with financial support from the Nederlandse Organisatie voor Wetenschappelijk Onderzoek (Netherlands Organization for Scientific Research, NWO) Grant number 16694. 
\end{acknowledgments}

\appendix
	\section*{Appendix A: DNS description}

\renewcommand{\theequation}{A\arabic{equation}}
\setcounter{equation}{0}

		In this appendix we briefly present the method used in the direct numerical simulation. We solve the Navier-Stokes equations (continuity and momentum equations) for incompressible flow,
		\begin{eqnarray}
		 \nabla \cdot \textbf{u} &=& 0, \label{eq:continuity} \\
			\rho\frac{\partial\textbf{u}}{\partial t} + \rho{\bm{\omega}} \times {\textbf{u}} + \nabla P &=& \rho\nu 		\Delta {\textbf{u}} + \rho{\textbf{a}}, \label{eq:N-S}
		\end{eqnarray}
		
	\noindent where $\textbf{u}$ is the velocity of the fluid, ${\bm{\omega}}=\nabla \times \textbf{u}$ is the vorticity, $P = p + \frac{1}{2} \rho\textbf{u}^{2}$, $\nu$ and $\rho$ are the kinematic viscosity and mass density of the fluid, $p$ is the periodic part of the static pressure. The term $\rho\textbf{a}$ corresponds to the mean pressure gradient and is the driving force density, which is uniform in space and in the streamwise direction, and chosen constant in time, in such a way that the Reynolds number based on the friction velocity is equal to 180.
	
\begin{figure*}[ht]
	\centering
	\includegraphics[width=\linewidth]{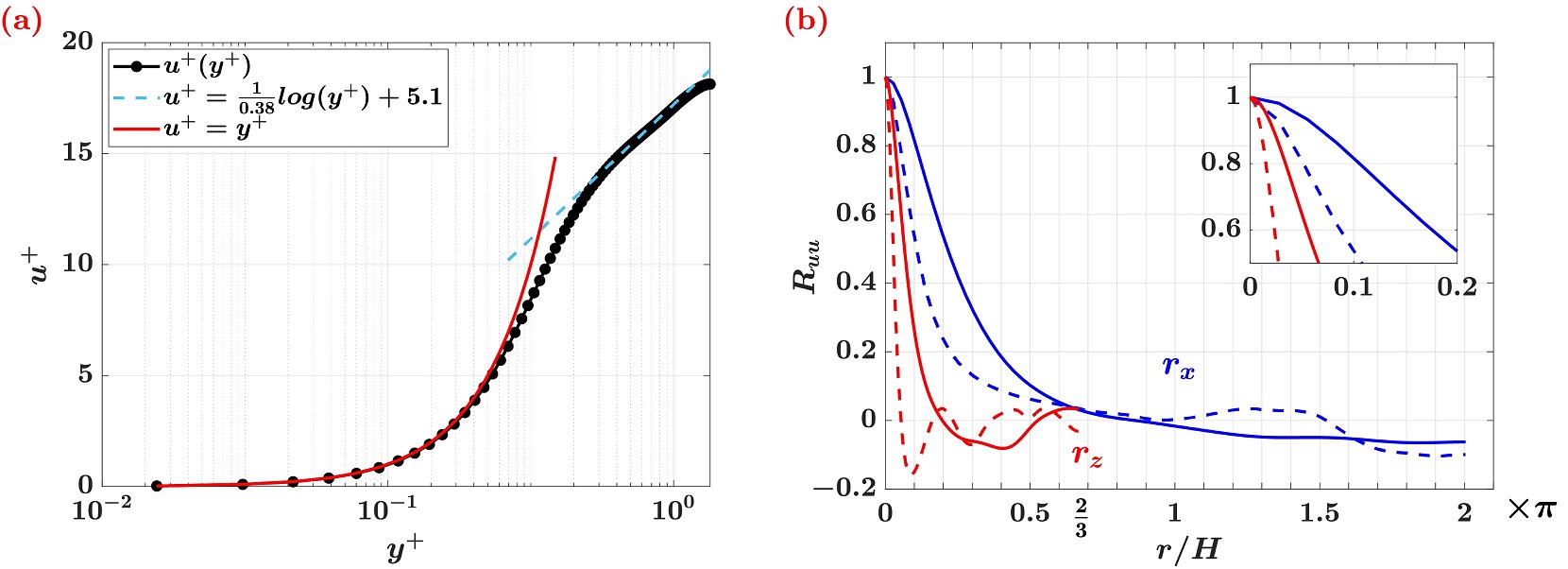}
	\caption{(a) Mean velocity profile (in wall units) as a function of the wall-normal coordinate, $y^+$. The law-of-the-wall is also shown. (b) Average two-points correlation, $R_{uu}$, of the streamwise velocity, $u$, at $y^+=3.5$ (solid lines) and at $y^+=180$ (dashed lines). $R_{uu}$ is plotted as a function of the spatial separations in the streamwise (blue) and spanwise directions (red), $r_x$ and $r_z$, respectively. \label{fig:Umean_autocorr}}
\end{figure*}

\begin{figure*}[ht]
	\centering
	\includegraphics[width=.98\linewidth]{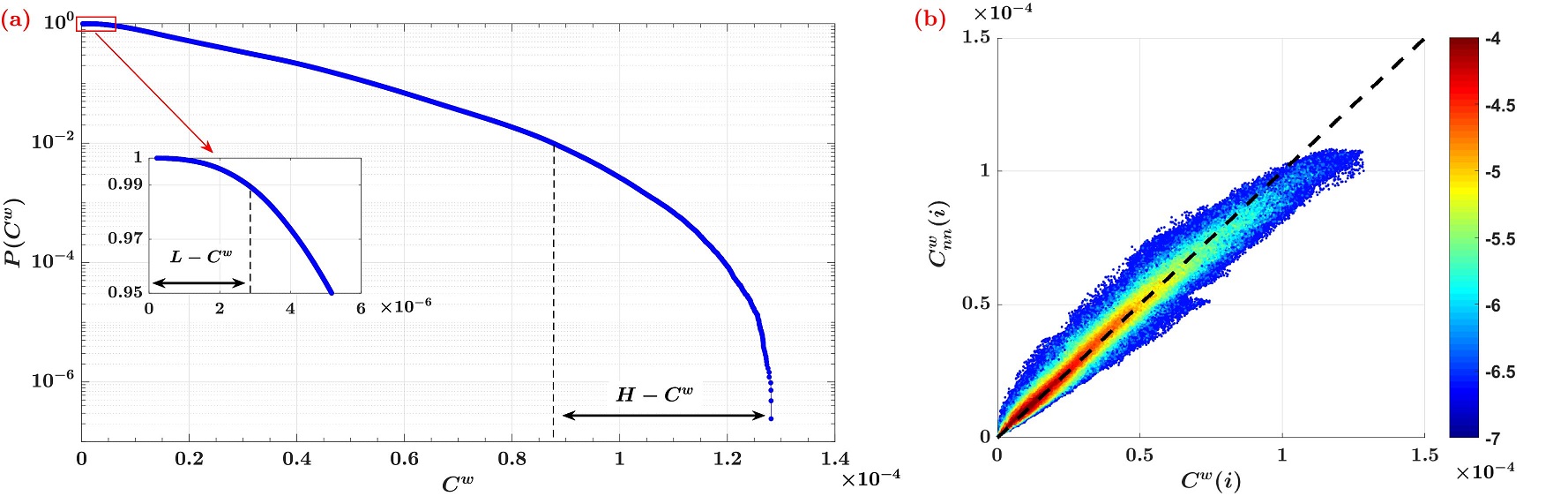}

	\caption{Global analysis of the network built on the wall-normal velocity component. (a) Cumulative $C^w$ distribution, $P(C^w)$. The inset is a zoomed view for small $C^w$ values, indicated by the red box. The ranges of $H-C^w$ and $L-C^w$ (shown in the inset) are highlighted. (b) Weighted average nearest neighbors assortativity measure, $C^w_{nn}(i)$, as a function of $C^w(i)$. Colors indicate the joint probability values (in $log_{10}$ scale) of variables $C^w(i)$ and $C^w_{nn}(i)$. The bisector is also displayed as a black dashed line. \label{fig:PVWC_assort_v}}
\end{figure*}

		The numerical method is based on the method used for DNS of turbulent channel flow by Kim et al.~\cite{kim1987turbulence}, but with the time integration method by Spalart et al. \cite{spalart1991}. In the two periodic directions a Fourier-Galerkin approach is used, whereas a Chebyshev-tau method is applied in the wall-normal direction. Instead of the velocity components, the wall-normal component of the vorticity vector and the Laplacian of the wall-normal velocity component are the dependent variables. In this way the incompressibility condition~(\ref{eq:continuity}) is automatically satisfied. The nonlinear terms in the Navier-Stokes equation~(\ref{eq:N-S}) are calculated in physical space by fast Fourier transform (FFT) with application of the 3/2 rule in both periodic directions. A combination of a three-stage second-order accurate Runge-Kutta method and the implicit Crank-Nicolson method is chosen according to~\cite{spalart1991}. This method has been used and validated extensively at frictional Reynolds numbers ranging between 150 and 950~\cite{geurts2012,kuerten2013,michalek2013,vreman2014a,vreman2014b}.

		For the present results the same grid S2 as in \cite{vreman2014a} has been chosen. This means that the domain size has a length $4\pi H$ in the streamwise direction $x$, $2H$ in the wall-normal direction $y$ and $\frac43 \pi H$ in the spanwise direction $z$, where $H$ denotes half the channel height. The number of Fourier modes in the streamwise direction equals 384, the number of grid points in the wall-normal direction equals 193 and the number of Fourier modes in the spanwise direction equals 192. This implies that in physical space the number of grid points equals $576\times193\times288$. The time step, $\Delta t$, used in the simulation equals $2.5\times10^{-4}H/u_{\tau}$, which implies that $\Delta t^+=\Delta t u_{\tau}^2/\nu=0.045$, where the superscript $+$ denotes wall units and $u_{\tau}$ is the frictional velocity. The number of time steps in the simulation is 5000, which corresponds to a time $T u_{\tau}/H=1.25$, or $T^+=225$ in wall units.

		Fig. \ref{fig:Umean_autocorr} shows the mean streamwise velocity profile and the average two-point spatial velocity correlations obtained from the DNS. The velocity profile (panel a) is compared with the law-of-the-wall and with $u^+=y^+$, which holds in the viscous sublayer, where $y^+<5$. The average two-point correlations, $R_{uu}$, shown in Fig. \ref{fig:Umean_autocorr}b are for the streamwise velocity component at two different wall-normal positions: $y^+=3.5$, very close to the wall, and $y^+=180$, in the center of the channel. Correlations are shown in both the streamwise and spanwise directions, and are calculated by averaging the correlation coefficients, $C_{i,j}(x_i,y_i,z_i;x_j,y_j,z_j)$, between pair of nodes $(i,j)$, along the homogeneous directions. They show that the extent of the domain in both periodic directions is sufficient for the average correlations to decay to zero.

\begin{figure}[h]
	\centering
	\includegraphics[width=\linewidth]{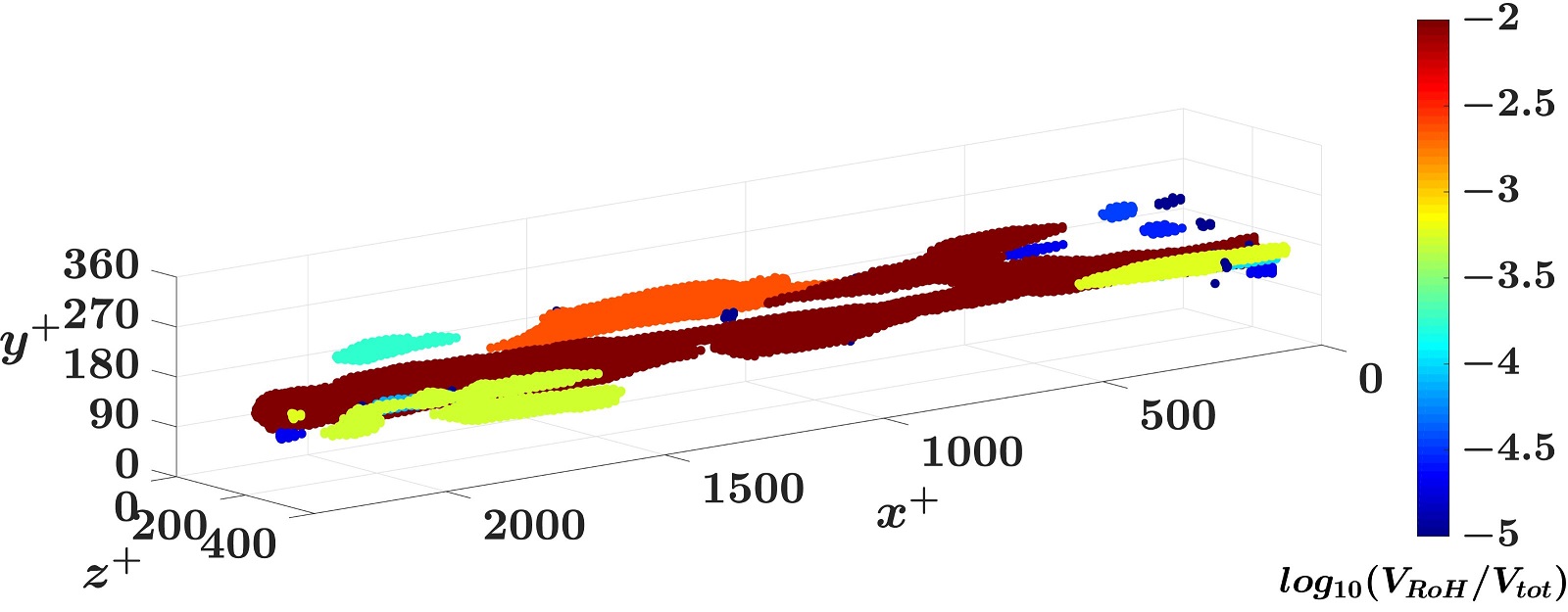}
	\caption{ 3D view of $H-C^w$ nodes, namely regions of hubs (RoHs) for the network built on wall-normal velocity. Color scale refers to the fraction of volume occupied by distinct RoHs, $V_{RoH}/V_{tot}$. \label{fig:apdx_RoH_V}}
\end{figure}

\begin{figure*}[ht]
	\centering
	\includegraphics[width=\linewidth]{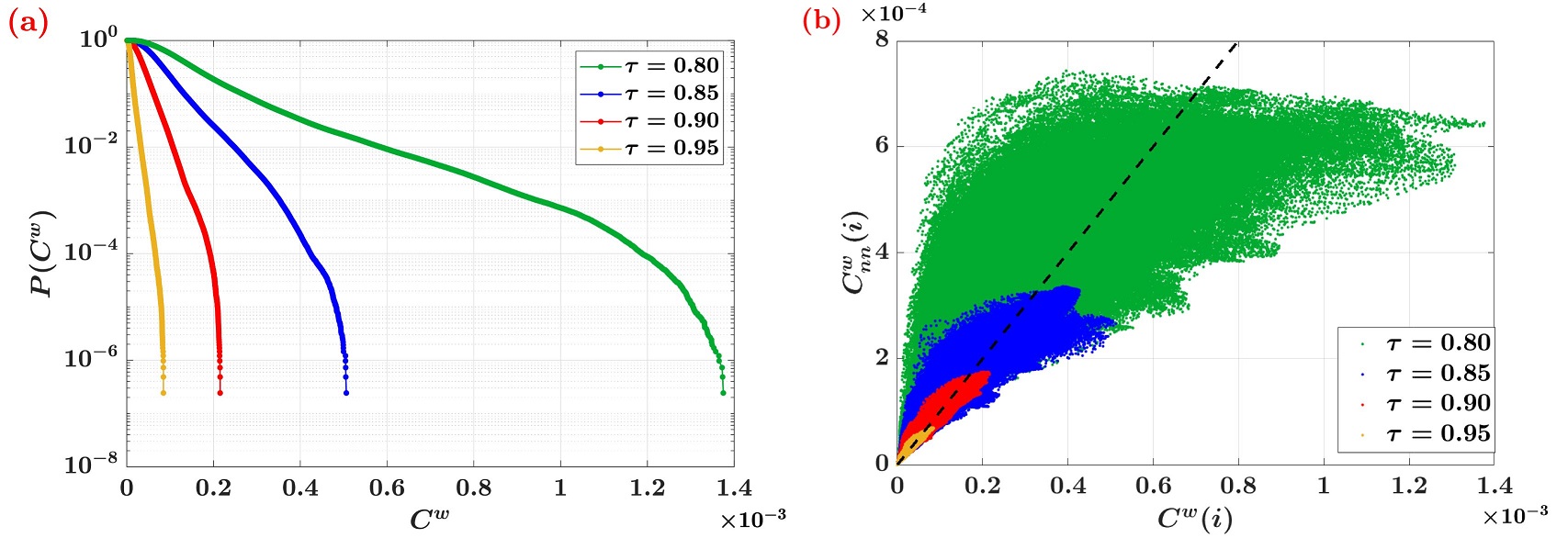}
	\caption{Network built on $u$. (a) Cumulative $C^w$ distribution, $P(C^w)$, and (b) weighted average nearest neighbors assortativity measure, $C^w_{nn}(i)$, as a function of $C^w$, for different thresholds $\tau$. \label{fig:apdx_P_VWC}}
\end{figure*}
	
	\section*{Appendix B: Network results for the $v$-component}	

	This appendix reports some results concerning the network based on the wall-normal velocity, $v$, with $\tau=0.85$, which are not included in the main text. They are shown here to give a comprehensive picture of this network.
	
	Fig. \ref{fig:PVWC_assort_v}a shows the $C^w$ cumulative probability, $P(C^w)$, while Fig. \ref{fig:PVWC_assort_v}b illustrates the average $C^w$ of neighbors of a generic node $i$, $C^w_{nn}(i)$. The network built on $v$ is strongly assortative, i.e., the nodes and their neighbors are close in space and share similar neighborhoods in terms of spatial extension (i.e., similar values of $C^w$). Therefore, the substantial absence of teleconnections makes the network of $v$ even more assortative than the network of $u$ (see also Fig. \ref{fig:apdx_tau_trends}a, in Appendix C). The $P(C^w)$ shown in Fig. \ref{fig:PVWC_assort_v}a sharply decays (if compared with the $P(C^w)$ of $u$ shown in Fig. \ref{fig:PVWC_assort_u}a), indicating that high values of $C^w$ are extremely rare in the network. Therefore, the global features of the network built on $v$ are similar to those of the network built on $u$, but more pronounced.

	The spatial location of the $H-C^w$ nodes of the network of $v$ is shown in a 3D view in Fig. \ref{fig:apdx_RoH_V}. As for the network of $u$, $H-C^w$ nodes are not scattered in the domain, but they tend to locally group into clusters elongated in the streamwise direction, as an effect of the streamwise advection. It should be noted that in this case, differently from the network of $u$, $H-C^w$ nodes only occur around the center of the channel. This happens because teleconnections enhance the centrality of nodes close to the walls, and this behaviour is magnified in the network based on $u$ rather than in the network on $v$.

	\section*{Appendix C: Parametric analysis}

\begin{figure*}[ht]
	\centering
	\includegraphics[width=\linewidth]{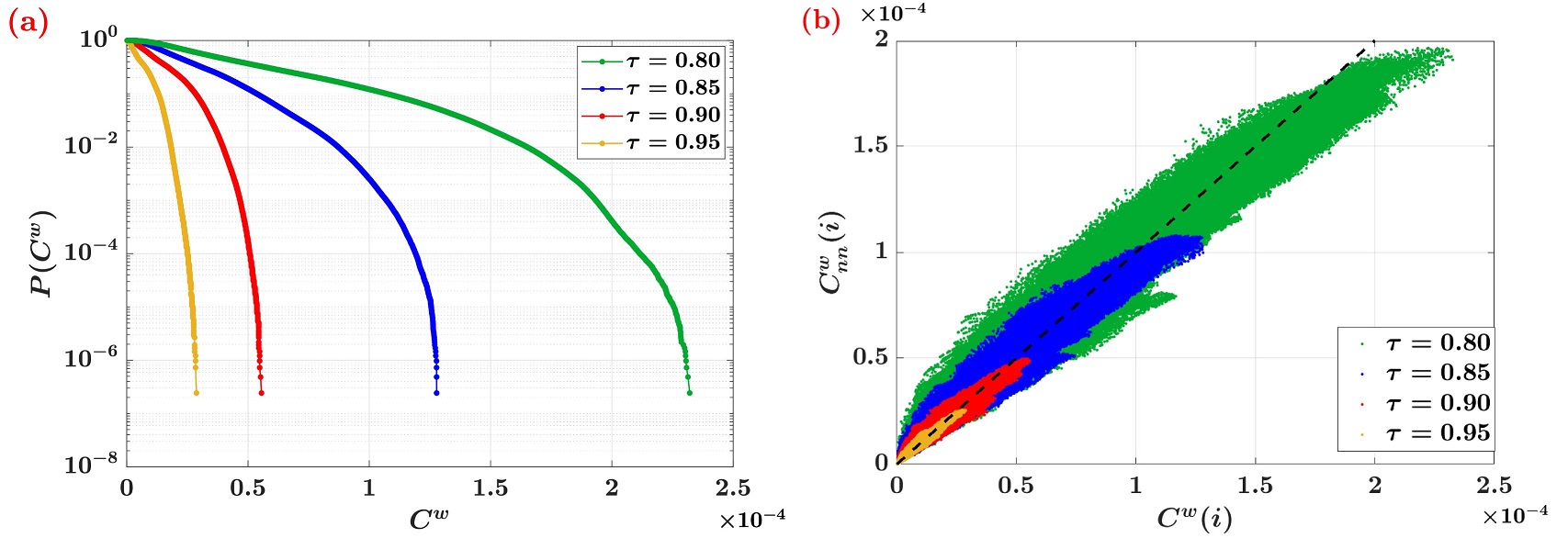}
	\caption{ Network built on $v$. (a) Cumulative $C^w$ distribution, $P(C^w)$, and (b) weighted average nearest neighbors assortativity measure, $C^w_{nn}(i)$, as a function of $C^w$, for different thresholds $\tau$. \label{fig:apdx_P_VWC_V}}
\end{figure*}

			In this section, a parametric analysis of the results is reported for different correlation thresholds, $\tau$. Besides $\tau = 0.85$, networks of both $u$ and $v$ velocity components for three different values were analyzed, namely $\tau = \left\lbrace0.8,0.9,0.95\right\rbrace$. Fig. \ref{fig:apdx_P_VWC}a shows the global scale results for the network of $u$: the cumulative $C^w$ distribution maintains a decreasing exponential behaviour for different values of $\tau$, with increasing slopes (in modulus) for increasing $\tau$. Since very high correlations are unlikely to appear, the overall $C^w$ values tend to decrease as $\tau$ increases. The same effect is shown in Fig. \ref{fig:apdx_P_VWC_V}a for the network built on $v$, with an increasing slope as $\tau$ increases. As for $\tau=0.85$, $H-C^w$ nodes for different thresholds are determined by considering the 99-th percentile of the corresponding $P(C^w)$ distribution.

			By focusing on the assortativity behaviour, we find that between $C^w_{nn}$ and $C^w$ an almost linear relation holds for the networks at different $\tau$, which is more evident for high values of $\tau$ (see Fig. \ref{fig:apdx_P_VWC}b and Fig. \ref{fig:apdx_P_VWC_V}b for the networks built on $u$ and $v$, respectively). This implies that the correlation-based networks always display assortative behaviours.

\begin{figure*}[ht]
	\centering
	\includegraphics[width=\linewidth]{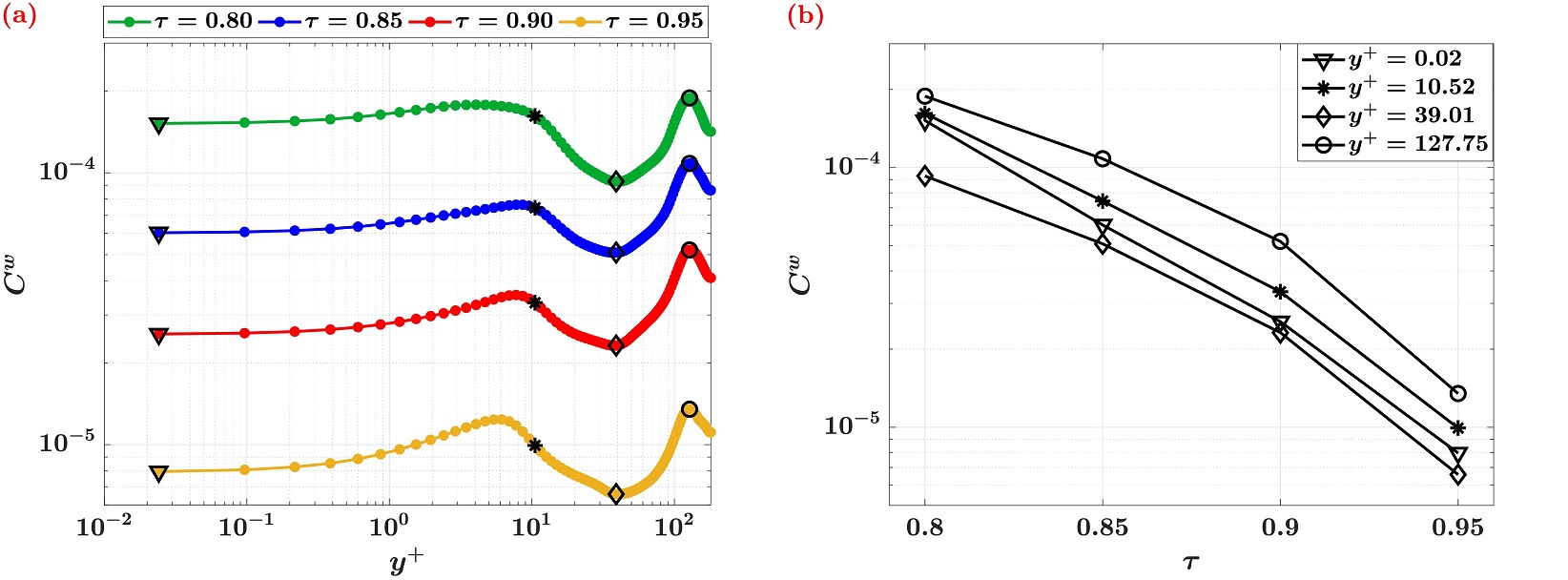}	
	\caption{Network built on $u$. (a) Volume-weighted connectivity, $C^w$, as a function of $y^+$ and averaged over the two homogeneous directions for different thresholds $\tau$. (b) $C^w$ trends at fixed $y^+$ coordinates (see panel (a)), as a function of $\tau$. \label{fig:apdx_y_avrg_VWC}}
\end{figure*}

\begin{figure*}[ht]
	\centering
	\includegraphics[width=\linewidth]{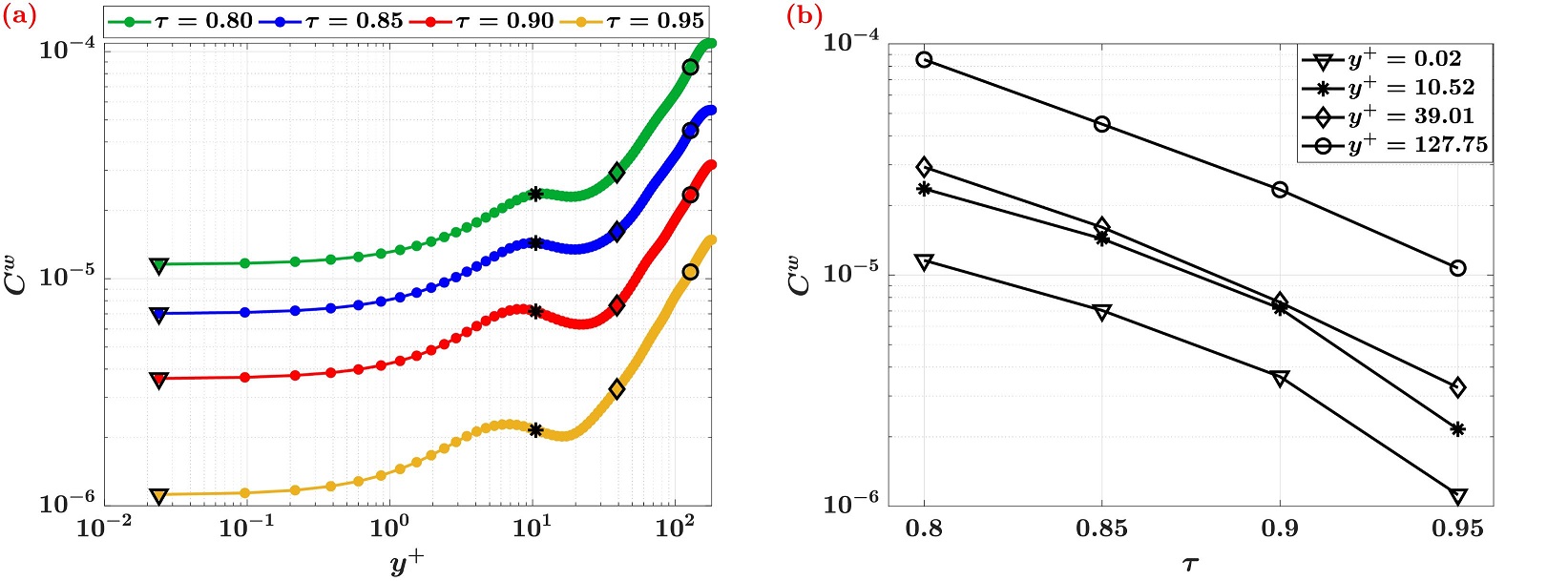}	
	\caption{Network built on $v$. (a) Volume-weighted connectivity, $C^w$, as a function of $y^+$ and averaged over the two homogeneous directions for different thresholds $\tau$. (b) $C^w$ trends at fixed $y^+$ coordinates (see panel (a)), as a function of $\tau$. \label{fig:apdx_y_avrg_VWC_V}}
\end{figure*}

			Moving to the meso-scale level of analysis, the average $C^w$ as a function of the wall-normal coordinate, $y^+$, is first investigated. As shown in Fig. \ref{fig:apdx_y_avrg_VWC}a (network built on $u$) and Fig. \ref{fig:apdx_y_avrg_VWC_V}a (network built on $v$), similar behaviours of average $C^w$ as a function of $y^+$ is found by changing $\tau$ and, in particular, the location of local peaks remains almost unchanged. In particular, we selected four representative $y^+$ locations and plotted the average values of $C^w$ as a function of $\tau$ in those wall-normal coordinates. As shown in Fig. \ref{fig:apdx_y_avrg_VWC}b (network built on $u$) and Fig. \ref{fig:apdx_y_avrg_VWC_V}b (network built on $v$), the trends of $C^w$ persist as $\tau$ changes, and they differ only by a constant value.
						
\begin{figure*}[ht]
	\centering
	\includegraphics[width=.95\linewidth]{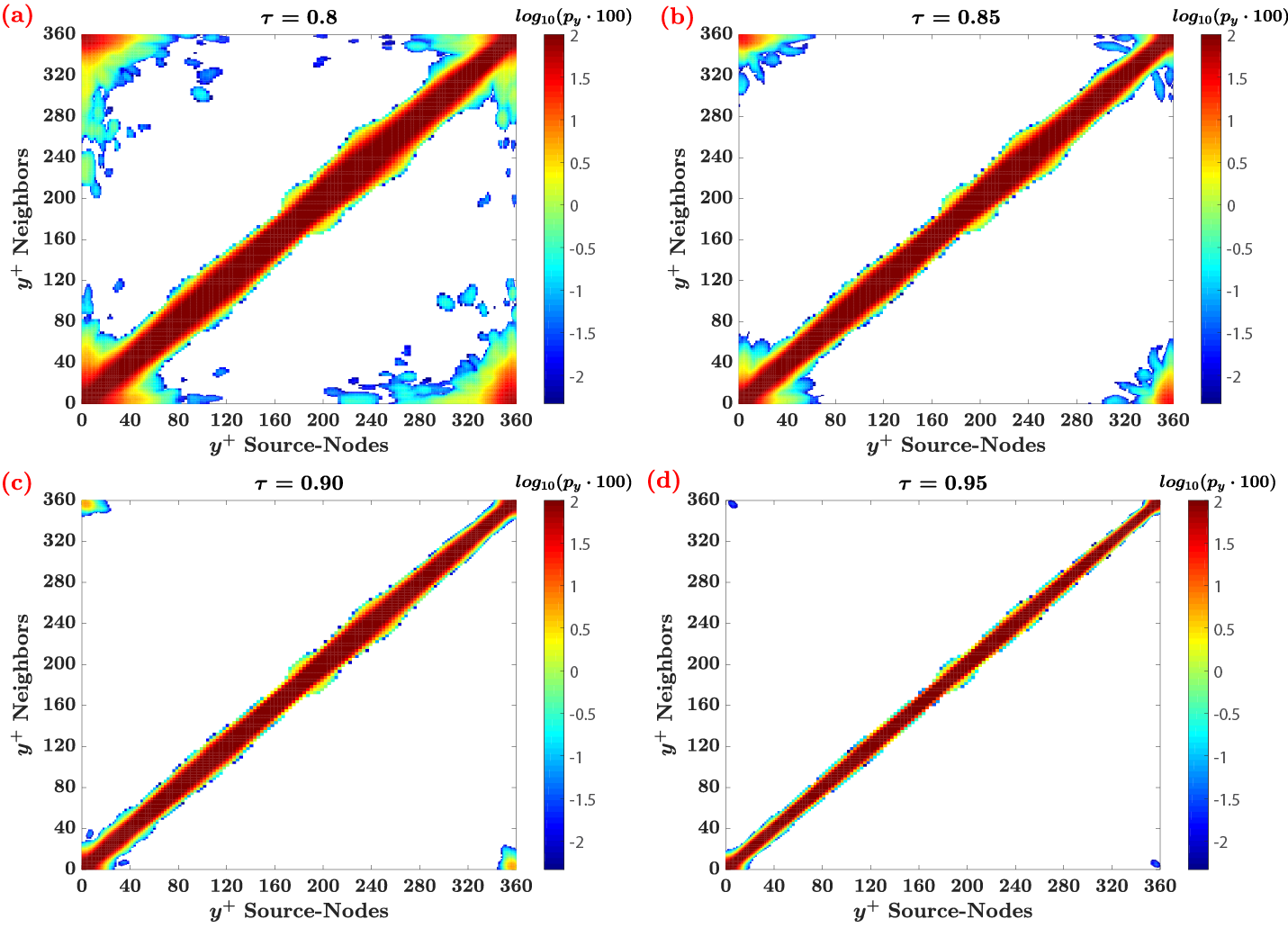}
	\caption{The probability (in log scale) that a source-node at a given $y^+$ is linked to a neighbor at another $y^+$ value, for network built with different threshold $\tau$. The color-scale ranges from $log_{10}(1/21600*100)\approx -2.34$ to $log_{10}(100)=2$, where $21600$ is the number of nodes in a plane at fixed $y^+$. \label{fig:apdx_Prob_link_y}}
\end{figure*}

		In order to investigate the presence of teleconnections between nodes close to both walls in the network of $u$, the probability that an arbitrary source-node at a fixed $y^+$ plane has a neighbor at another $y^+$ value is shown in Fig. \ref{fig:apdx_Prob_link_y} for different $\tau$ values. By increasing the correlation threshold, nodes at any $y^+$ tend to have first neighbors closer to them in the wall-normal direction (the diagonal part of the plot stretches), but intra- and inter-wall teleconnections are still present for $\tau=0.95$ for source-nodes very close to the wall (see Fig. \ref{fig:apdx_Prob_link_y}d).

\begin{figure*}[ht]
	\centering
	\includegraphics[width=\linewidth]{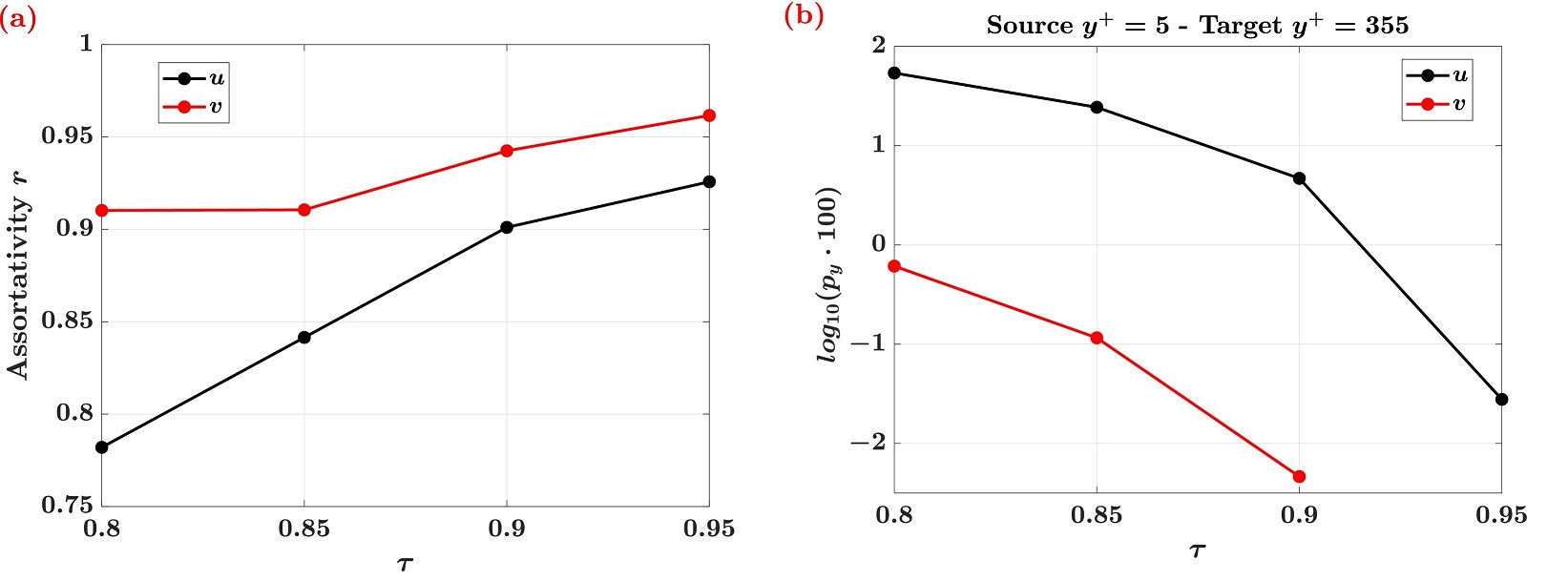}	
	\caption{(a) Pearson assortativity coefficient as a function of the threshold, $\tau$. (b) The probability that a source-node at $y^+=5$ is linked to a neighbor at $y^+=355$, for the networks built with different threshold $\tau$ (see also Fig. \ref{fig:apdx_Prob_link_y} for the network based on $u$). \label{fig:apdx_tau_trends}}
\end{figure*}

 	By combining the results shown so far, we can conclude that the overall topological features of the networks do not substantially change, but they re-scale for different values of $\tau$. In order to have a more comprehensive overview, trends of network features as a function of $\tau$ are shown in Fig. \ref{fig:apdx_tau_trends}. First, the assortativity coefficient, $r$, is reported in Fig. \ref{fig:apdx_tau_trends}a. The assortativity coefficient is the Pearson correlation coefficient of the degree centrality of pairs of neighbors, thus giving a scalar indication of the assortativity of the network \cite{costa2007characterization}. The network is assortative/disassortative if $r$ is positive/negative. High positive values of $r(\tau)$ are found, confirming the outcome of the plots of $C^w_{nn}$ as a function of $C^w$. Moreover, Fig. \ref{fig:apdx_tau_trends}b shows the probability that a source-node at $y^+=5$ is linked to a neighbor at $y^+=355$, for different threshold $\tau$. It clearly emerges that, in the network of $u$ (black line) there is a higher probability to find wall-wall teleconnections than in the network of $v$ (red line). However, the trends of the probability are very similar (note that the probability for $v$ is zero for $\tau=0.95$), showing again that the networks built on the $u$ and $v$ similarly change with $\tau$.

\begin{figure*}[ht]
	\centering
	\includegraphics[width=.95\linewidth]{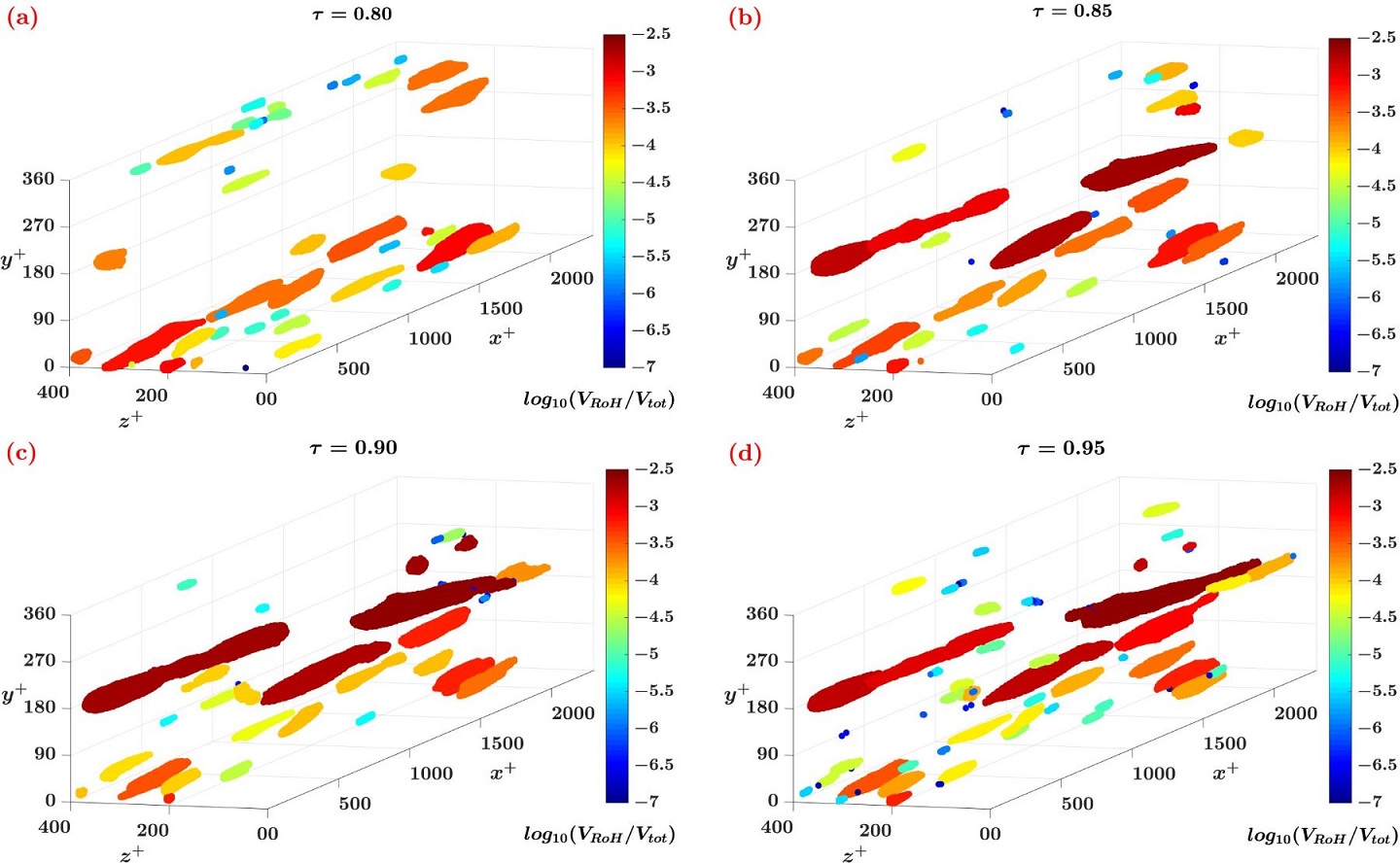}	
	\caption{Network built on $u$. 3D views of $H-C^w$ nodes, namely regions of hubs, RoH for different threshold values, $\tau$. Color scale refers to the fraction of volume occupied by distinct RoHs, $V_{RoH}/V_{tot}$. \label{fig:apdx_degree_3D}}
\end{figure*}
		
		Finally, the spatial location of the $H-C^w$ nodes of the network built on $u$ for $\tau= \left\lbrace 0.8, 0.85, 0.9, 0.95\right\rbrace$ is shown in Fig. \ref{fig:apdx_degree_3D}. As $\tau$ is changed, $H-C^w$ nodes close to the wall preserve their overall spatial organization, i.e. they group into $x$-elongated spatially-connected regions, namely RoHs. The scenario is also similar for $H-C^w$ nodes close to the center of the channel for $\tau \geq 0.85$. For $\tau=0.8$, instead, nodes far from the walls tend to have lower $C^w$ values than nodes close to the walls, because the latter ones are involved in a large number of teleconnections that markedly increases their centrality in the network. This causes a smaller number of RoHs in the channel center. The same conclusion also holds for the network built on $v$.
			
			We can conclude that, for sufficiently high correlation thresholds, the effect of the mean flow is maintained and wall-wall teleconnections occur. Therefore, in the range of $\tau$ values investigated, the main features of the network built at $\tau=0.85$ are also found, but the specific values of the metrics clearly change, because of different edge density values as $\tau$ changes. A much lower threshold value would certainly have the effect to include more long-range links in the networks, but links would lose physical significance.
					
\bibliography{biblio_channel_Re180_ACC}

\end{document}